\documentclass[12pt]{article}

\title{The Anomalous Nambu-Goldstone Theorem in Relativistic/Nonrelativistic Quantum Field Theory}
\author{Tadafumi Ohsaku}

\date{today}

\begin{document}

\maketitle


\begin{flushleft}

{\it In der Welt habt ihr Angst; aber seid getrost, ich habe die Welt \"{u}berwunden.} ( Johannes, Kapitel 16 )

{\it It always seems impossible until it is done.} ( Nelson Mandela )

\vspace{5mm}

{\bf This Paper is Dedicated for Our Brave Fighters and Super-Heroes for the Fundamental Human Rights Around the World.}

\end{flushleft}

\vspace{5mm}

{\bf Abstract}: 
The anomalous Nambu-Goldstone ( NG ) theorem which is found 
as a violation of counting law of the number of NG bosons of the standard ( normal ) NG theorem
in nonrelativistic and Lorentz-symmetry-violated relativistic theories 
is studied in detail, with emphasis on its mathematical aspect from Lie algebras, geometry to number theory.
The basis of counting law of NG bosons in the anomalous NG theorem is examined by Lie algebras ( local ) and Lie groups ( global ).
A quasi-Heisenberg algebra is found generically in various symmetry breaking schema of the anomalous NG theorem,
and it indicates that it causes a violation/modification of the Heisenberg uncertainty relation in an NG sector
which can be experimentally confirmed.
This fact implies that we might need a framework "beyond" quantum mechanical apparatus to describe quantum fluctuations 
in the phenomena of the anomalous NG theorem
which might affect formations of orderings in quantum critical phenomena.
The formalism of effective potential is presented for understanding the mechanism of anomalous NG theorem with the aid of our result of Lie algebras.
After an investigation on a bosonic kaon condensation model with a finite chemical potential as an explicit Lorentz-symmetry-breaking parameter,
a model Lagrangian approach on the anomalous NG theorem is given for our general discussion.
Not only the condition of the counting law of true NG bosons, 
but also the mechanism to generate a mass of massive NG boson is also found by our examination on the kaon condensation model. 
Furthermore, the generation of a massive mode in the NG sector is understood by the quantum uncertainty relation of the Heisenberg algebra,
obtained from a symmetry breaking of a Lie algebra, which realizes in the effective potential of the kaon condensation model.
Hence the relation between a symmetry breaking scheme, a Heisenberg algebra, a mode-mode coupling, and the mechanism of mass generation
in an NG sector is established. 
Finally, some relations between the Riemann hypothesis and the anomalous NG theorem are presented.

\vspace{5mm}

KEYWORDS: Normal and anomalous Nambu-Goldstone theorem, Lie algebras and Lie groups, Heisenberg algebra, 
the Heisenberg uncertainty relation, quantum fluctuation and phase transition, 
differential geometry, number theory, the Riemann hypothesis, spin systems, QCD. 

\vspace{5mm}

\section{Introduction}

This paper belongs to the recent efforts to intend to make a final answer on the controversial ( or, not well-organized ) 
issue on the counting law of the number of Nambu-Goldstone ( NG ) bosons/modes
in a nonrelativistic and/or Lorentz-symmetry-broken system. 
We mainly concentrate on the cases of so-called continuous internal symmetries expressed by Lie groups, 
commute with the spacetime 4-translations and the Lorentz 4-rotations.
We restrict ourselves on the cases of four-dimensional spacetime:
Two- and three-dimensional cases are outside of our discussion due to the Mermin-Wagner-Coleman theorem~[11,48,52].
The Nambu-Goldstone theorem ( established around 1960-1962~[26,27,60,61] ) states that a zero-mass/gap particle 
( Nambu-Goldstone boson ) naturally arises from a theory associated with a spontaneously-broken-symmetry generator, 
with one-to-one correspondences between broken generators and the NG bosons.
Here we call this standard situation as the "normal" NG theorem.
It is a famous fact that, in an anomalous case, 
the number of NG bosons does not coincide with the number of broken generators of ${\rm Lie}(G)$ in a case of nonrelativistic field theory
( $G$ is a Lie group which gives a symmetry of the system, Lie$(G)$ denotes its corresponding Lie algebra ).
Nielsen and Chadha found this phenomena in 1976~[64]. 
After those discoveries in fundamental physics, almost four decades, there is no precise solution 
or any explanation of reasoning of the Nielsen-Chadha anomaly. 
Quite recently ( 2002-2013 ), this controversy has been started to be resolved by several works in theoretical physics.

In this section, we summarize those recent results of their works appeared in literature. 
To clarify our discussions and perspectives, 
one should employ a classification:
Lorentz = symmetric, spontaneously broken, explicitly broken;
Lie group = symmetric, spontaneously broken, explicitly broken.
Thus totally $3\times 3=9$ cases we have in general.
In those cases, we consider here some examples where the Lorentz symmetry is explicitly broken,
while a Lie group symmetry is spontaneously/explicitly broken.
We will show that these two examples are understood by a single framework.
In principle, our anomalous NG theorem is fall into the category of "spontaneous" cases without an explicit symmetry breaking. 
Other cases remain for our further study in future.
In fact, recent several works given in literature also consider the cases that the Lorentz symmetry is explicitly broken 
or a nonrelativistic case,
while a Lie group of internal symmetry is spontaneously broken.

First, we discuss the main result of Refs.[8,34,81,82,83,84] which is summarized into the following inequality 
for the number of NG bosons given by Watanabe and Brauner: They gave it without any derivation~[81]:
\begin{eqnarray}
n_{NG} &\ge& n_{BS} -\frac{1}{2}{\rm rank}\langle 0| [Q^{A},Q^{B}] |0\rangle.
\end{eqnarray}
Here, $n_{NG}$ gives the number of NG bosons they would be observed ( when a norm is positive and physical ),
$n_{BS}$ is the number of broken generators of ${\rm Lie}(G)\simeq T_{e}G$
( $G$; a Lie group which describes a symmetry of system we consider, $e$; the origin ),
$Q^{A}$ and $Q^{B}$ ( $A,B = 1, \cdots, n_{BS}$ ) imply the conserved charges of broken generators, and $|0\rangle$ is the vacuum of a theory. 
It should be mentioned that the rank is defined for a matrix with matrix entries of indices $AB$.
While, we know the very simple and universal law that the number of broken generators $n_{BS}$ equals
the sum of the number of "true" NG bosons and the number of "massive" NG bosons:
\begin{eqnarray}
n_{true-NG} &=& n_{BS} - n_{massive-NG}.
\end{eqnarray}  
It is clear for us that $\frac{1}{2}{\rm rank}\langle[Q^{A},Q^{B}]\rangle$ must give the number of massive modes 
( the rank of Hessian of effective action expanded by NG modes around a VEV, see the paper of S. Weinberg in our references~[86] ):
The matrix is given in a quadratic form of broken generators.
Thus, our issue is now understood how to find a law in which the broken generators contain massive modes.
Let us compare the anomalous behavior with a case of explicit symmetry breaking.
The paper of Weinberg~[86] discusses a chiral Lagrangian with an explicit symmetry breaking mass term proportional to 
\begin{eqnarray}
{\rm tr}(e^{iQ}\Phi^{-iQ}) &=& {\rm tr}(g\Phi g^{-1}) = {\rm tr}{\rm Ad}(G(\Phi)),
\end{eqnarray}
( $\Phi$; some bosonic fields ) and gives a formula of square of mass parameters of NG bosons by the second-order derivative of the mass term.
It is given by the bracket like the following form, namely, a term given by twice actions of adjoints:
\begin{eqnarray}
[[\Phi,Q^{A}],Q^{B}]
\end{eqnarray}
and if we regard this as a matrix of indices $(A,B)$, then its dimension is exactly equal with $[Q^{A},Q^{B}]$ appeared in the formula
for counting the number of NG bosons.
Therefore, there is a similarity between an anomalous and an explicit symmetry breaking at the Lie algebra level {\it beside the factor 1/2}.
Later, we will see how this factor 1/2 arises in our anomalous NG theorem.
If a theory has $n_{ex}$ explicit symmetry breaking mass parameters, and if 
\begin{eqnarray}
n_{ex} \ge \frac{1}{2}{\rm rank}\langle[Q^{A},Q^{B}]\rangle,
\end{eqnarray} 
then 
\begin{eqnarray}
n_{NG} = n_{BS} -n_{ex}
\end{eqnarray}
may holds, since an explicit symmetry breaking parameter enforces that an NG boson always has a finite mass.

While, after the work of Watanabe and Brauner, 
Hidaka~[34] derived the equation ( replaces $\ge$ to $=$ in the above inequality (1) ) via the generalized Langevin formalism:
His formalism is essentially the same with the method of effective action.
Thus, later we utilize the effective action formalism, both Lorentz-violating relativistic and nonrelativistic cases.

Note that in a Poincar\'{e} invariant theory, a charge $Q$ is Lorentz scalar when it is conserved, $[P^{\mu},Q]=0$.
This is due to the Coleman-Mandula no-go theorem~[12].
Schaefer et al. pointed out in their paper~[77]
that if $Q$ is given from the zeroth-component of conserved vector current of an internal symmetry, 
then it cannot have a nonvanishing VEV for a Lorentz-symmetric vacuum.
They argue that we need $\langle Q\rangle\ne 0$ for realizing an anomalous behavior of NG theorem. 
We need a careful discussion on it.
In fact, we now consider a theory of Lorentz-violating system,
hence we cannot restrict $Q$ as a Lorentz scalar.
Moreover, we should distinguish the cases of Wigner phase and Nambu-Goldstone ( NG ) phase.
In a Wigner phase, $Q|0\rangle =0$ is immediately concluded since $Q$ is a conserved quantity, while $Q|0\rangle \ne 0$ in an NG phase,
and it does not look like a conserved quantity: This is the essential part of the NG theorem.
The symmetry of a Lagrangian ( theory ) and its vacuum do not coincide with each other in an NG phase.
Due to the unitary inequivalence, one can not definitely say about what $e^{iQ}|0\rangle$ gives.
For example, the vector $Q|0\rangle$ cannot be normalized in an NG phase.
It might be possible to say that $|0\rangle$ or $Q|0\rangle$ are not $G$-modules in the naive sense.
While, if a theory spontaneously breaks its vacuum symmetry, and if the Lorentz symmetry is broken under a certain manner,
then we lost the basis of the statement of $\langle 0|Q|0 \rangle=0$ even though it will be defined 
as an integral of three-dimensional total volume/space:
The physical situation of those VEVs may be determined self-consistently.
We emphasize the fact that this discussion is valid for a quantum theory but not for classical systems,
since we take a VEV of a quantum operator.
We will also give some insights on the case where a broken charge is a Lorentz-violating generic tensor.
Hence we have another classification:
Vacuum = Lorentz symmetric/Lorentz violated, N\"{o}ther charge = Lorentz symmetric/Lorentz violated.

According to our short observation of several previous results in literature,
we classify NG bosons into,
(i) true NG bosons as exactly massless particles, 
(ii) massive NG bosons in our anomalous NG theorem, 
(iii) pseudo NG bosons which have finite masses due to explicit symmetry breaking parameters in the Lagrangian of the beginning. 
Several examples of (iii) have been studied, for example, in Refs.~[16,70].
Our terminology presented here is not the same with the famous classification of NG bosons 
given by S. Weinberg for Lorentz-invariant relativistic cases~[86]:
Our present discussion should be understood as a generalization/extension of it.

Our several classifications are summarized into the table given in the next page:
\begin{table}
\begin{tabular}{|c|c|}
\hline  
Theory (Lagrangian) & LS, explicitly LV \\    
\hline 
Vacuum & LS, spontaneously LV, explicitly LV  \\   
\hline 
Lie group & symmetric, SB, AB, EB \\   
\hline 
Discrete ( C, P, T ) & symmetric, SB, EB  \\    \hline  
N\"{o}ther charge & LS, LV   \\      \hline
NG boson & true, anomalously massive, pseudo    \\   \hline 
\end{tabular}
\end{table}
Here, several abbreviations mean: 
LS = Lorentz symmetric, 
LV = Lorentz violated, 
SB = spontaneously broken, 
AB = anomalously broken, 
EB = explicitly broken.
"Discrete" indicates a discrete symmetry, typically as a charge conjugation ( C ), a parity ( P ), and a time-reversal ( T ).

This paper is organized as follows: 
In sec. 2, several typical symmetry breaking schema will be studied from their Lie algebra/group aspects,
and will find several characteristic features of them in our anomalous NG theorem, which never takes place in the standard NG case. 
In sec. 3, an effective potential formalism will be investigate to understand the mechanism of our anomalous NG theorem
by employing our Lie-algebra results.
In sec. 4, a kaon condensation model with a finite chemical potential will be examined to obtain our viewpoint
on a generic Lagrangian of NG bosons which gives the phenomenon of anomalous NG theorem.
Then we will construct a generic Lagrangian which cause the anomalous NG theorem, in sec. 5.
Some relations between our anomalous NG theorem and number theory, especially the Riemann zeta function, will be presented in sec. 6.
Finally, the concluding remarks will be given in sec. 7.

\section{Lie Algebras, Lie Groups, and Symmetry Breaking Schema}

In this section, our anomalous NG theorem is examined by Lie algebras ( give some local characters of NG sectors )
and Lie groups ( contain informations on global aspects/structures of NG manifolds ),
with employing several examples.

First, we would like to pay attention on the following fact before obtaining a general discussion of symmetry breakings.
In a breaking scheme of a symmetry, it is not always the case that a Lie group $G$ is broken to a Lie subgroup $H$ to give a coset $G/H$:
Thus, an examination on cosets as results of symmetry breakings is not enough for studying the ( global ) nature of 
( normal/explicit+dynamical/anomalous ) NG theorem.  
For example, let us consider some examples of $SO(3)$ or $SU(2)$. 
( You can compare with the case of $U(2)$, or the electroweak symmetry breaking of the Standard Model! )
The Hamiltonian of spin systems of ferro- and antiferromagnets may belong to $SO(3)$ and sometimes $SU(2)$~[87] 
( they are locally isomorphic, ${\rm Lie}SO(3)\simeq{\rm Lie}SU(2)\simeq{\rm Lie}USP(2)$,
and thus one has a freedom to choose one of them at least at the Lie algebra level ).
The isospin space also be described by $SU(2)$.
In a ferromagnetic case, quite a lot of works consider broken generators as $s_{1}$ and $s_{2}$ of $SU(2)$ while $s_{3}$ remains "unbroken."
( $s_{a}$, $a=1,2,3$ $\in{\rm Lie}(SU(2))$, by using the representation of Pauli matrices. )
This breaking scheme is schematically denoted as $SU(2)\to U(1)$, but it does not give a coset:
This breaking scheme does not have a coset ( quotient ) topology,
since the set of $g_{3}=e^{i\theta\sigma_{3}}$ does not form a closed normal subgroup.
In this case, two massless NG bosons may be expected but we find only one due to the Nielsen-Chadha anomaly.

Let $G$ be a Lie group which gives the symmetry of a system,
and its Lie algebra as ${\bf g}={\rm Lie}(G)$.
Let $S^{\alpha}$ ( $\alpha=1,\cdots, {\rm dim}(G)-n_{SB}$ ) denote the generators ( a set of bases of ${\rm Lie}(G)$ ) 
correspond to remaining symmetries,
and let $X^{\beta}$ ( $\beta=1,\cdots, n_{SB}$ ) imply the broken generators.
From the orthogonality condition of the bases of ${\bf g}={\rm Lie}(G)$,
the Lie brackets $[S^{\alpha},X^{\beta}]$ always belong to the linear space of broken generators.
While, any commutator of broken generators will be given by a linear combination of all generators,
\begin{eqnarray}
[X^{\beta},X^{\gamma}] = \sum c^{\alpha}S^{\alpha} + \sum c^{\delta}X^{\delta}.
\end{eqnarray}  
Hence, if the corresponding charges $Q^{S^{\alpha}}$ of $S^{\alpha}$ are conserved and simultaneously they are Lorentz symmetric,
and if the vacuum of the theory is also Lorentz symmetric, then $Q^{S^{\alpha}}|0\rangle=0$ is concluded immediately.
On the contrary, $Q^{X^{\delta}}|0\rangle\ne 0$ ( for $\forall \delta$ ) even if they are Lorentz scalar.
If the breaking scheme $G\to H$ ( $G$, $H$; Lie groups ) forms a coset $G/H$ and if it is a symmetric space,
then any Lie bracket of broken generators belongs to ${\rm Lie}(H)$~[20,32,43]:
\begin{eqnarray}
[X^{\alpha},X^{\beta}] \subset {\bf h} = {\rm Lie}(H), \quad S^{\alpha} \in {\bf h}, \quad X^{\beta} \in {\bf m}, \quad {\rm Lie}(G) = {\bf g} = {\bf h} + {\bf m}.
\end{eqnarray}  
In that case, the VEV of any $[X^{\alpha},X^{\beta}]$ always vanishes in the case of Lorentz symmetric conserved charges belong to ${\bf h}$.
( Therefore, if the relation of Watanabe and Brauer is correct and is valid also in a symmetric space, 
then the number of massive modes is given as a function of VEVs of symmetric generators. ) 
It is interesting for us to consider several models defined over Riemannian ( global ) symmetric spaces of the Cartan classification~[32].
Later, we will discuss how a local nature of anomalous NG theorem is extended to a global structure in a symmetric space.

The equation of Lie algebra
$[X^{\beta},X^{\gamma}] = \sum c^{\alpha}S^{\alpha} + \sum c^{\delta}X^{\delta}$
means that the left hand side of commutator is expanded by the linear form of the right hand side:
Namely, this formula counts the dimension of the linear space which the commutator belongs to,
and especially after taking a VEV of both side of this equation, it gives a subspace which the VEV of commutator is described,
such like a two-dimensional surface in a three-dimensional space. 
In this sense, the VEV of this equation is "algebro-geometric." 
After employing a method of compactification suitable for a breaking scheme,
a deformation theory and moduli space for such an algebraic variety~[30,31] could be introduced ( but not always ).
Especially in the symmetric space mentioned above,
${\rm dim}[X^{\alpha},X^{\beta}] = {\rm dim}{\rm Lie}(H)$.
If the matrix $\langle 0|[X^{\alpha},X^{\beta}]|0\rangle$ is obtained from the second-order derivative of an effective potential 
expanded by NG bosons associated with broken generators,  
the matrix might contain a nonvanishing part, embedded in the total part of the second-order derivative, 
which gives the finite mass eigenvalues for the NG bosons.
Namely, the dimension of the matrix of nonvanishing part is the dimension of the linear space of massive NG bosons. 
( In fact, the proof given in the paper of Watanabe and Murayama, Ref.~[82], can be interpreted as a calculation of basis set of 
the linear space which the mass matrix of NG bosons belongs. )
Any type of non-vanishing VEV of $\sum c^{\alpha}S^{\alpha} + \sum c^{\delta}X^{\delta}$ 
defines which pair $[X^{\alpha},X^{\beta}]$ forms a non-vanishing matrix element.
You should notice that the rank of a matrix implies the dimension of a linearly-independent components of a matrix.
It should be mentioned that the dimension of mass matrix is obtained after taking a VEV of the vacuum $|0\rangle$, 
and thus at the moment one can say nothing about the mass matrix when one takes a displacement from the vacuum $|0\rangle$
( a mass matrix is given by displacements of displacements ).
It also should be examined how these conditions of counting the dimension of massive modes in a mass matrix defined in the linear space
of Lie algebra has the validity, in a Higgs-type bosonic model, 
an effective action of composite model like the Nambu$-$Jona-Lasinio ( NJL ) model or QCD, 
or in the case of Coleman-Weinberg mechanism~[13]. 
In this paper, we mainly consider a Goldstone-Higgs-type bosonic Lagrangian/theory to investigate our anomalous NG theorem.
Since a displacement caused by a Lie group action to an order parameter
( where it is a composite or an elementary field ) is given by an adjoint action of a Lie group,
thus, our result should be valid also in an NJL-type composite model.
This is, of course, also the case in a Coleman-Weinberg mechanism of symmetry breaking.
Later, we examine a Goldstone-Higgs type bosonic model of kaon condensation,
while an examination of NJL or QCD demands us further investigation as another paper, due to the fact that
those theories demand us some heavy calculations.

A lot of discussions use an $SU(2)$ model with two broken generators $\sigma_{x}$, $\sigma_{y}$ while $\sigma_{z}$ is symmetric
( $(\sigma_{x},\sigma_{y},\sigma_{z})\in{\rm Lie}(SU(2))$ ).
In this case, $Q^{z}\propto \sigma_{z}$ can be regarded as an unbroken charge even though this breaking scheme does not give a coset, 
\begin{eqnarray} 
[Q^{x},Q^{y}] = iQ^{z} \to {\rm symmetric}
\end{eqnarray}
and thus the commutator always vanishes for a Lorentz-symmetric vacuum if $Q^{z}$ is Lorentz-scalar and a conserved quantity.
Note that only a symmetric generator ( namely $Q^{z}$ ) appears in the right hand side in this expansion, 
even though the breaking scheme does not give a coset ( quotient ) and of course not a symmetric space: 
This breaking scheme is special from several points.
Moreover, if the following relations of VEVs holds, 
\begin{eqnarray}
& & \langle[Q^{x},Q^{y}]\rangle = i\langle Q^{z}\rangle \ne 0,  \\
& & \langle[Q^{x},Q^{z}]\rangle = -i\langle Q^{y}\rangle = 0,  \\
& & \langle[Q^{y},Q^{z}]\rangle = i\langle Q^{x}\rangle = 0, 
\end{eqnarray}
then, they are isomorphic with the three-dimensional Heisenberg algebra,
\begin{eqnarray}
[x,y] = z, \qquad [x,z] = 0, \qquad [y,z] = 0.
\end{eqnarray}
Thus, the insight of Nambu given in Ref.~[62] which states that $Q^{x}$ and $Q^{y}$ form a canonical conjugate pair 
in the case of ferromagnet $\langle Q^{z}\rangle\ne 0$ is mathematically natural.
Since a vacuum of the theory must be chosen to evaluate VEVs of these brackets,
the theory of $SU(2)$ is expanded at the origin of the Lie group manifold by the NG bosonic coordinates.
Thus, the transformation from the Lie algebra to the Heisenberg algebra is achieved at the origin
of the Lie group and the corresponding Heisenberg group.
Physically, such a Heisenberg-algebra relation directly concludes the Heisenberg uncertainty principle in the "dynamical degrees of freedom",
hence two NG-bosonic coordinates generated by the conserved charges over a group manifold may acquire a quantum uncertainty.
A comment on the appearance of a symplectic vector space or group might be possible as some literature already have done ( Refs.~[81-84] ), 
though the notion and structure of (quasi-)Heisenberg algebras/groups
are better to emphasize the quantum nature, since a Poisson bracket of classical mechanics also satisfies a symplectic structure.
In fact, a Heisenberg algebra is a central extension of an algebra of symplectic vector space,
and an automorphism of Heisenberg algebra is given by a group of symplectic type.
Such a symplectic vector space of course defines a symplectic structure such like $\omega=\sum dp_{i}\wedge dq_{i}$ up to an isomorphism, 
which is given by a so-called Lagrangian subspace/submanifold.
It should be emphasized that the transform from a Lie algebra to a Heisenberg algebra under the prescription given above
is not achieved by some kind of perturbation or an analytic expansion ( such as a deformation quantization of Poisson manifold~[45] )
but by a functor, a functorial manner provided by quantum field theory.
It is a known fact that both a three-dimensional Heisenberg group and $SU(2)$ can be embedded into $SU(2,1)$~[46].
Thus, both the three-dimensional Heisenberg algebra and ${\rm Lie}(SU(2))$ can be derived from the same Lie group,
and it is interesting for us to know how those "submanifolds" are related with each other inside a larger Lie group,
and how an automorphism of ${\rm Lie}(SU(2))$ are related with that of the Heisenberg algebra, vice versa.
( The groups of automorphisms of $G$ and ${\rm Lie}(G)$ are isomorphic in general.
Thus, the transform from ${\rm Lie}(SU(2))$ to the three-dimensional Heisenberg algebra may give a global correspondence
via their automorphism groups. )
Let us investigate this problem by ourselves. 
${\rm Lie}(SU(2,1))$ is an eight-dimensional algebra, 
while ${\rm Lie}(SU(2))\simeq {\rm Lie}(SU(1,1))\simeq {\rm Lie}(Sp(2))\simeq {\rm Lie}(SL(2,{\bf R}))\simeq {\rm Lie}(SL(2,{\bf C}))$ 
and the three-dimensional Heisenberg algebra define three-dimensional linear spaces.
Thus, the three-dimensional spaces of ${\rm Lie}(SU(2))$ and the Heisenberg algebra 
are embedded in the eight-dimensional space of ${\rm Lie}(SU(2,1))$,
and the anomalous NG theorem gives a mapping ( possibly a bijection ) between two spaces. 
$SU(2)$ defines a sphere $S^{2}$, namely a curve or a compact Riemann surface, and thus the corresponding three-dimensional Heisenberg group
should also define a curve or a Riemann surface.
Hence, the correspondence ( functor ) between ${\rm Lie}(SU(2))$ and the three-dimensional Heisenberg algebra cause
a correspondence between two curves or Riemann surfaces with some globalization of those Lie algebras,
especially via exponential mappings.
The NG bosons of the breaking scheme $SU(2)\to U(1)$ define a subset of $S^{2}$ ( a set of circles ).
This implies that the NG bosons of this breaking scheme gives a subspace of a Riemann surface,
and thus the three-dimensional Heisenberg algebra also gives a subspace of a Riemann surface as the Heisenberg group manifold.

Let us examine the case of $SU(3)$ ( its Lie algebra is isomorphic with Lie$SL(3,{\bf C})$ ). 
The definition of the Gell-Mann matrix representation of ${\rm Lie}(SU(3))$ is
\begin{eqnarray}
& & [Q^{A},Q^{B}] = if^{ABC}Q^{C}, \quad Q^{C} \in {\rm Lie}(SU(3)), \quad (A,B,C=1,2,\cdots,8)    \nonumber \\
& & f^{123} = 1, \quad f^{147} = f^{165} = f^{246} = f^{257} = f^{345} = f^{376} = \frac{1}{2}, \quad f^{458} = f^{678} = \frac{\sqrt{3}}{2}.
\end{eqnarray}
For example, in the case of $\langle Q^{3}\rangle\ne 0$, $\langle Q^{8}\rangle \ne 0$ with all of other generators have vanishing VEVs, 
the set of following VEVs gives a "quasi" Heisenberg algebra:
\begin{eqnarray}
& & \langle[Q^{1},Q^{2}]\rangle = i\langle Q^{3}\rangle,  \quad  \langle[Q^{1},Q^{3}]\rangle = \langle[Q^{2},Q^{3}]\rangle = 0,   \nonumber \\
& & \langle[Q^{4},Q^{5}]\rangle = \frac{i}{2}\langle Q^{3}\rangle + \frac{i\sqrt{3}}{2}\langle Q^{8}\rangle,  \nonumber \\
& & \langle[Q^{4},Q^{3}]\rangle = \langle[Q^{5},Q^{3}]\rangle = \langle[Q^{4},Q^{8}]\rangle = \langle[Q^{5},Q^{8}]\rangle = 0,   \nonumber \\
& & \langle[Q^{6},Q^{7}]\rangle = -\frac{i}{2}\langle Q^{3}\rangle + \frac{i\sqrt{3}}{2}\langle Q^{8}\rangle,  \nonumber \\
& & \langle[Q^{6},Q^{3}]\rangle = \langle[Q^{7},Q^{3}]\rangle = \langle[Q^{6},Q^{8}]\rangle = \langle[Q^{7},Q^{8}]\rangle  = 0.
\end{eqnarray} 
The VEVs of all other brackets vanish and "commute."
Strictly speaking, the set of VEVs in this case does not give a Heisenberg algebra in the sense of its definition given below, 
and we need to remove $Q^{3}$ or $Q^{8}$ from the algebra to set $\langle Q^{3}\rangle =0$ or by $\langle Q^{8}\rangle =0$. 
While we observe that a pairwise decoupling takes place. 
All of $(Q^{1},Q^{2},Q^{4},Q^{5},Q^{6},Q^{7})$ are broken at the case $\langle Q^{3}\rangle \ne 0$ and $\langle Q^{8}\rangle \ne 0$,
or at the case $\langle Q^{3}\rangle \ne 0$ and $\langle Q^{8}\rangle =0$ ( those cases give the scheme $SU(3)\to U(1)\otimes U(1)$ ),
while $(Q^{4},Q^{5},Q^{6},Q^{7})$ are broken and $(Q^{1},Q^{2},Q^{3},Q^{8})$ remain unbroken 
( this case gives the breaking scheme $SU(3)\to SU(2)\otimes U(1)$ ) in the case $\langle Q^{3}\rangle = 0$ and $\langle Q^{8}\rangle \ne 0$.
However, if we change the representation of Lie$SU(3)$ from the Gell-Mann-type to others such as canonical basis,
then we find all generators except the Cartan subalgebra will be broken 
when we give a finite VEV for one of elements of Cartan subalgebra of Lie$SU(3)$.
Since a physical phenomenon which depends on our choice of representation of a Lie algebra never takes place in the nature,
we conclude that the choice $\langle Q^{3}\rangle = 0$ and $\langle Q^{8}\rangle \ne 0$ is a special case of the Gell-Mann representation,
never occurs in the nature.   
Moreover, when we seek a Heisenberg algebra in a Goldstone-type bosonic model of $SU(3)$, 
we need to choose the form of VEV to make these charges broken when $\langle Q^{3}\rangle\ne 0$ and $\langle Q^{8}\rangle =0$
( the case $SU(3)\to SU(2)\otimes U(1)$ mentioned above ).
In such a case, the bosonic field $\Phi$ belongs to ${\bf 3}$-representation, and $\Phi^{\dagger}Q^{3}\Phi\ne 0$ and $\Phi^{\dagger}Q^{8}\Phi=0$
gives an additional condition to the three components of $\Phi$.
Namely, we have to perform a variation of a subspace of complex three ( real six ) dimensional space: This is not natural.
Thus, we conclude that the diagonal breaking of $SU(3)$ always gives not a Heisenberg but a quasi-Heisenberg algebra.
This fact indicates that the Heisenberg uncertainty relation might be modified in quantum mechanical description of fluctuations 
( namely, NG bosons )
of the NG sector of the diagonal breaking of $SU(3)$, such that,
\begin{eqnarray}
\Delta \chi^{1}\Delta\chi^{2} \ge C^{a}, \quad \Delta\chi^{4}\Delta\chi^{5}\ge C^{b}, \cdots.
\end{eqnarray}
Hence we speculate the deviation from the Heisenberg algebra given by a set of VEVs of the Cartan subalgebra measures
which pair of NG modes is more "classical" and which pair of NG modes has a quantum fluctuation stronger than others.
A deviation from the Heisenberg-type uncertainty relation might affect on quantum fluctuation in quantum phase transition.

The definition of Heisenberg algebra is
\begin{eqnarray}
[p_{i},q_{j}] = \delta_{ij}z, \quad [p_{i},z] = [q_{j},z] = 0,
\end{eqnarray}
where $(p_{1},\cdots,p_{n},q_{1},\cdots,q_{n},z)$ gives the generator of the algebra.
Thus, the number of generators must be odd in the Heisenberg algebra.
Here, $z$ is a central element of the Heisenberg algebra.
Hence the finite VEVs of elements of Cartan subalgebra give the center of the ( quasi ) Heisenberg algebra we have obtained, 
namely a central extension.  
A quasi-Heisenberg algebra is defined to be
\begin{eqnarray}
[p_{i},q_{j}] = \delta_{ij} \sum_{\alpha}z_{\alpha}, \quad [p_{i},z_{\alpha}] = [q_{j},z_{\alpha}] = 0.
\end{eqnarray}
Moreover, the expansion
\begin{eqnarray}
\langle[X^{\beta},X^{\gamma}]\rangle &=& \sum c^{\alpha}\langle S^{\alpha}\rangle + \sum c^{\delta}\langle X^{\delta}\rangle
\end{eqnarray}  
must be pairwise decoupled to obtain a Heisenberg algebra: This is in general not the case.
Hence, to obtain a Heisenberg algebra via VEVs of Lie brackets of generators of a Lie algebra,
a subset of generators of odd number must give a subalgebra.
In this sense, an algebra generically obtained from the VEVs of conserved charges should be called as a deformed/quasi Heisenberg algebra.
The uncertainty relation in a quasi-Heisenberg algebra should be investigated in detail, 
since it might give a violation or a modification of the ordinary Heisenberg uncertainty relation 
in a spontaneous symmetry breaking system, and it would be confirmed experimentally
from physical behaviors of NG sectors in a condensed matter or a nucleus, especially in their quantum critical phenomena~[10].

Let us examine how a quasi-Heisenberg algebra arises by using the general theory of Cartan decomposition, 
Cartan matrix, and canonical basis in the case of semisimple Lie algebra Lie$(G)$~[20,32,43]. 
In this case, via the root system ( so-called Cartan-Weyl basis ),
\begin{eqnarray}
{\bf g} &=& {\bf h}\oplus\bigoplus_{\lambda\in R}{\bf g}_{\lambda},  \\
{\bf g}_{\lambda} &=& \bigl\{ a\in {\bf g}: [h_{j},a] = \lambda(h_{j})a, \quad \forall h_{j} \in {\bf h}\bigr\},
\end{eqnarray} 
( $\lambda(h_{j})$: Cartan matrix, $R$: roots ),
the Lie algebra is generically defined by
\begin{eqnarray}
& & {\bf g} = {\bf h}\oplus {\bf e}\oplus {\bf f}, \quad h_{i}, h_{j} \in {\bf h}, \quad e_{i}, e_{j} \in {\bf e}, \quad f_{j} \in {\bf f}, \\
& & [h_{i},h_{j}] = 0,  \\
& & [e_{i},f_{j}] = \delta_{ij}h_{i},  \\
& & [h_{i},e_{j}] = a_{ij}e_{j},  \\
& & [h_{i},f_{j}] = -a_{ij}f_{j}.
\end{eqnarray}
Here, $a_{ij}$ denote the Cartan matrix, and ${\bf h}$ is the Cartan subalgebra.
Thus, if $\langle e_{i}\rangle = \langle f_{j}\rangle =0$ ( $\forall i,j$ ) while some of the bases of Cartan subalgebra take finite VEVs, 
$\langle h_{j}\rangle\ne 0$, then a pairwise decoupling takes place and a quasi-Heisenberg algebra is embedded in the total algebra.  
A Heisenberg pair is given by the algebra basis of a positive and a negative roots.
Thus, we obtain the following theorem:
\begin{flushleft}
{\bf Theorem}:
{\it Let $G$ be a semisimple Lie group and let Lie$(G)$ be its semisimple Lie algebra.
Let us assume the case where a theory only has VEVs of generators toward the directions of Cartan subalgebra.
Then the Cartan subgroup remains unbroken, and the generators of Lie algebra will be pairwisely decoupled by taking their VEVs,
they form a quasi-Heisenberg algebra.
This type of decoupling never takes place in a Lorentz-invariant system
due to the vanishing condition $\langle Q \rangle =0$ of a conserved charge.}
\end{flushleft}
In such a situation of this theorem, a Lagrangian of NG bosons may be pairwise decomposed inside the linear space of NG bosons
at least in the quadratic part of the Lagrangian of NG boson fields 
( we discuss how and when such a decomposition takes place in the NG sector of a theory in sec. 5 ).
In the case of breaking scheme $G\to$ Cartan subgroup, where all group elements except the Cartan subgroup are broken,
and all generators of Cartan subalgebra take non-vanishing VEVs,
then one can count the number of pairs ( namely, the number of mode-mode couplings of NG bosons ) which give a quasi-Heisenberg algebra:
\begin{eqnarray}
n_{pair} &=& \frac{1}{2}\Bigl[ {\rm dim}{\rm Lie}(G)-{\rm rank}{\rm Lie}(G) \Bigr].
\end{eqnarray}
Note that the rank of Lie$(G)$ coincides with the dimension of Cartan subalgebra.
If $G\to H$ gives a symmetric space $G/H$, then 
\begin{eqnarray}
{\rm rank}{\rm Lie}(G) &=& {\rm dim}{\rm Lie}(H) = {\rm dim}[X^{\alpha},X^{\beta}]
\end{eqnarray}
holds. Thus,
\begin{eqnarray}
n^{symmetric-space}_{pair} &=& \frac{1}{2}\Bigl[ {\rm dim}{\rm Lie}(G)-{\rm dim}{\rm Lie}(H) \Bigr]    \nonumber \\
&=& \frac{1}{2}\Bigl[ {\rm dim}{\rm Lie}(G)-{\rm dim}[X^{\alpha},X^{\beta}] \Bigr].
\end{eqnarray} 
Moreover, if the Lagrangian of any pair of NG bosons is given in the form 
as only one mode of a pair is massive by the mixing of modes inside the pair
( this situation will be given by an NG boson Lagrangian in sec. 5 ),
then the number of massive NG bosons equals the number of pairs, and
\begin{eqnarray}
n_{BS} &=& n_{NG} + \frac{n_{pair}}{2} = n_{NG} + \frac{1}{2}\Bigl[ {\rm dim}{\rm Lie}(G)-{\rm rank}{\rm Lie}(G) \Bigr].
\end{eqnarray} 
Since $[X^{\alpha},X^{\beta}]=\sum c^{\gamma}S^{\gamma}+\sum c^{\delta}X^{\delta}$,
\begin{eqnarray}
{\rm dim}\langle[X^{\alpha},X^{\beta}]\rangle = {\rm dim}\Bigl( \sum c^{\gamma}\langle S^{\gamma}\rangle \Bigr) = {\rm rank}{\rm Lie}(G)
\end{eqnarray}
holds in the diagonal breaking case. 
Hence we get the following equation for a diagonal breaking:
\begin{eqnarray}
n_{BS} &=& n_{NG} + \frac{1}{2}\Bigl[ {\rm dim}{\rm Lie}(G)-{\rm dim}\langle [X^{\alpha},X^{\beta}]\rangle \Bigr].
\end{eqnarray}
Here, ${\rm dim}{\rm Lie}(G)-{\rm dim}\langle [X^{\alpha},X^{\beta}]\rangle$ gives the number of Heisenberg pairs in the diagonal breaking scheme.
Since $n_{BS}=$dimLie$(G)-$dimLie$(H)$, we get
\begin{eqnarray}
n_{NG} &=& \frac{1}{2}\Bigl[ {\rm dim}{\rm Lie}(G)-{\rm rank}{\rm Lie}(G) \Bigr].
\end{eqnarray}

More general case is examined by utilizing (20)-(26).
When $\langle h\rangle\ne 0$, $\langle e\rangle = \langle f\rangle =0$ ( a diagonal breaking ),
then $e$ and $f$ are broken since $[\Phi,e]\ne 0$, $[\Phi, f]\ne 0$, $\Phi\in h$.
If $\langle h\rangle\ne 0$, $\langle e\rangle\ne 0$, $\langle f\rangle =0$, 
then $h$, $e$, $f$ are broken since $[\Phi, h]\ne 0$, $[\Phi,e]\ne 0$, $[\Phi, f]\ne 0$, $\Phi\in h\oplus e$.
In the latter case, the algebra of VEVs is not a (quasi) Heisenberg-type. 
An investigation on the case of all generators are broken, $G\to$ nothing, becomes complicated to give a general theory.

Since a Heisenberg group and its Lie algebra are realized on a symplectic vector space,
one can introduce a Darboux basis of a symplectic vector space, corresponds to the canonical coordinates,
to express the Heisenberg algebra.
Then the Heisenberg algebra acquires a geometric implication.
Moreover, one can introduce an operator algebra analysis of the anomalous NG theorem via 
the Stone-von Neumann theorem~[75]. 
Therefore, a unification of algebra, analysis, and geometry takes place in our anomalous NG theorem. 
The Heisenberg algebra $(p,q,z)$ obtained from our $SU(2)$ model is a special example of,
\begin{eqnarray}
X \stackrel{p}{\rightarrow} Y \stackrel{q}{\rightarrow} Z, \quad
X \stackrel{q}{\rightarrow} Y' \stackrel{p}{\rightarrow} Z', \quad 
Z \ne Z'.
\end{eqnarray}
Here, $X,Y,Z,Y',Z'$ implies some mathematical sets, and we regard the canonical pair $(p,q)$ is given by certain types of morphisms.
The center $z$ given by the VEV $\langle Q^{z}\rangle$ in the $SU(2)\to U(1)$ case measures how $Z$ and $Z'$ are different.
This kind of noncommutativity appears in algebras of monodromy, holonomy, etc.
This is the geometric nature of the Heisenberg algebra as the essence of quantum mechanics.
From the aspect of Heisenberg groups, our Heisenberg algebra coming from the anomalous NG theorem of an $SU(2)$ model 
can be related with
the three-dimensional compact Iwasawa manifold obtained from $\Gamma\backslash G/H$ of the three-dimensional Heisenberg group $G$, 
where
\begin{eqnarray}
G &=& \left(
\begin{array}{ccc}
1 & a & b \\
0 & 1 & c \\
0 & 0 & 1 
\end{array}
\right),  \quad a,b,c \in {\bf R}, \\
H &=& \{e\},  \\
\Gamma &=& G \cap GL(3,{\bf Z}).
\end{eqnarray}
It is a known fact that a complex structure is found in a three-dimensional Iwasawa manifold~[40].
From the Kodaira-Spencer theory of deformation of complex structure of a complex manifold~[44], 
we know the obstruction of deformation of a complex structure is determined by the cohomology group of the manifold,
namely it is one of global aspects/structures of the NG manifold of our anomalous NG theorem.

An interesting fact is that a three-dimensional Heisenberg algebra $[p,q]=z$, $[p,z]=[q,z]=0$ is constructed
by the differential operators such as $p=\partial_{1}-\frac{x^{2}}{2}\partial_{3}$,
$q=\partial_{2}+\frac{x^{1}}{2}\partial_{3}$, $z=\partial_{3}$:
In that case, $(p,q,z)$ forms an orthogonal frame of an appropriate manifold.
It is interesting for us to compare this fact with the representation of differential operator expression of the $sl_{2}$-triple.
It is a well-known fact that a Heisenberg algebra can be expressed by a Weyl algebra,
\begin{eqnarray}
[x_{i},\partial_{j}] = -\delta_{ij}, \quad [x_{i},x_{j}] = [\partial_{i},\partial_{j}] = 0, 
\end{eqnarray}
and the Weyl-algebra expression of our quasi-Heisenberg algebra gives us a further implication
of mathematical and geometric nature of our Lorentz-violating NG boson Lagrangian ( see sec. 5 ).
A Weyl algebra is a simple N\"{o}therian integral domain, and it has a global dimension $n$.
It should be mentioned that any term higher than the second-order of the expansion of an adjoint action expressed by an exponential mapping, 
a similarity transformation $g^{-1}Qg$ ( $Q$; a conserved charge ), 
or in a Killing form ( kinetic term ) of the Lagrangian or the effective potential, 
are fall into the fundamental relation of quasi-Heisenberg algebra (18) after taking their VEVs, 
and thus those expansions are effectively "terminated" at the quasi-Heisenberg algebra
( such a truncation can take place, of course, in a quantum theory ), 
and thus, only the subset of (quasi-)Weyl algebra of universal enveloping algebra appears in a theory:
This case is coming from the fact that our theoretical framework is suitable in the vicinity of the ground state
of the system determined by a choice of the form of VEVs
and we consider a Heisenberg algebra, not a Heisenberg group.
The Weyl algebra itself is isomorphic with the Moyal algebra of deformation quantization, thus the global character 
of our theory of anomalous NG theorem will acquire a connection with the Moyal-Weyl deformation quantization.
Moreover, a Weyl algebra defines several differential operators which directly connects with theory of $D$-modules~[4].
A Weyl algebra of an infinite order, possibly isomorphic with a deformation quantization, will be entered into 
our anomalous NG theorem when we consider the corresponding Heisenberg group:
Naively, the Weyl algebra is terminated at the order of the elementary relations ( Lie brackets, (38) ) of the corresponding quasi-Heisenberg algebra,
as we have stated,
while a linear transformation of a basis of representation space of the quasi-Heisenberg algebra caused by an operation of Heisenberg group 
( an adjoint action to the Heisenberg algebra ) gives the Weyl algebra which can acquire its higher-order
products and derivatives of algebras expanded by the set $(x_{j},\partial_{j})$. 
To make this matter consistent, we need to obtain the notion of quasi-Heisenberg group.    
It should be investigated that a Moyal-Weyl type deformation quantization for the quasi-Heisenberg/quasi-Weyl algebra
starting from a Poisson manifold defined by a Poisson structure of broken generators,
which might give some results confirmed by experiments.
From our observation of $SU(3)$, we can propose the following modified Moyal-Weyl product:
\begin{eqnarray}
f*g &=& f \exp \Bigg( (\sum_{j}\langle h_{j}\rangle)\sum_{A,B}
\frac{\overleftarrow{\partial}}{\partial\chi^{A}}\frac{\overrightarrow{\partial}}{\partial\chi^{B}}-\frac{\overleftarrow{\partial}}{\partial\chi^{B}}\frac{\overrightarrow{\partial}}{\partial\chi^{A}} \Bigg) g   \nonumber \\
&=& fg + \sum_{j}\langle h_{j}\rangle \{f,g\}_{PB} + \cdots,    \\  
\{f,g\}_{PB} &=& \frac{\partial f}{\partial\chi^{A}}\frac{\partial g}{\partial\chi^{B}}-\frac{\partial f}{\partial\chi^{B}}\frac{\partial g}{\partial\chi^{A}} 
\end{eqnarray}
This modified Moyal-Weyl product provides an example of a multi-Planck-"constant" model which has been proposed by the author in Ref.~[70].
The set of VEVs of the Cartan subalgebra generators gives several Planck "constants" which depend on a position of 
the Lie group manifold or a local geometry, mainly curvatures of $V_{eff}$.
( An example of deformation quantization of Heisenberg manifolds, see ~[73]. )
It should be mentioned that our approach to the anomalous NG theorem heavily depends on Lie algebras of conserved charges
and they are the special cases of current algebras, for example, 
$[j^{A}(x_{0},{\bf x}),j^{B}(x_{0},{\bf y})]=if^{ABC}\delta^{(3)}({\bf x}-{\bf y})j^{C}(x)$.
Hence, our result may be reformulated by those current algebras and their globalizations ( it may be called as "current groups",
and they might introduce new mathematics in our NG theorem, especially from the context of integral geometry and operator algebras ).
In Ref~[46], a quasiconformal mapping of a Heisenberg group is studied.
While, a quasiconformal mapping is obtained in a Moyal-Weyl deformation quatization~[67].
It is interesting for us to unify these approaches.

In more generic case, a Lie algebra is defined as a direct sum of finite number of Lie subalgebras:
\begin{eqnarray}
{\bf g} &=& {\bf g}_{1} + \cdots + {\bf g}_{l},
\end{eqnarray}
and each of them has a root space decomposition. 
Thus the theorem given above can be stated differently:
\begin{flushleft}
{\bf Theorem}:
{\it Any diagonal breaking which remains the Cartan subalgebra unbroken converts the Lie algebra into a direct sum of a finite number of 
quasi-Heisenberg algebras and Abelian algebras, via taking the VEVs of algebra generators.
Turn to the group theory, such a breaking scheme gives a direct product of quasi-Heisenberg groups and Abelian Lie groups
via the anomalous NG theorem in quantum field theory.
The quasi-Heisenberg group and Abelianized group act on the effective action/potential of the theory
and its low energy effective theory.}
\end{flushleft}

Since this paper considers internal symmetries mainly for our anomalous NG theorem, 
we choose a semisimple Lie group such as $SU(n)$, $SO(n)$ and $Sp(n)$ as the main subject,
though we can examine more generic cases of $SL(n,F)$ and $GL(n,F)$
( $F$: a number field of characteristic zero ).
Note that any Lie group has an associated Lie algebra. 
Moreover, especially ${\rm Lie}(SL(n,F))$ and ${\rm Lie}(GL(n,F))$ are finite dimensional,
we have a functor which gives corresponding simply-connected Lie groups.
Hence our analysis presented here is valid to those Lie algebras.
The Ado-Iwasawa theorem states that a finite-dimensional Lie algebra defined over a field $F$ has a faithful finite-dimensional representation.
A general consideration of some cases of ${\rm Lie}(SL(n,F))$ and ${\rm Lie}(GL(n,F))$ contains theories of $SU(N)$, $SO(N)$, ..., as
their special examples.
Especially, $GL(2,{\bf R})$ is disconnected into two parts according to the signature $\pm 1$ of determinant,
thus a problem of covering on the space of NG bosons, a problem of the global nature of the NG theorem 
similar to the case of gauge orbits, 
gives an interesting subject which has a strong connection with number theory.
Though, for examining the mechanism to generate massive NG bosons, group elements in the vicinity of identity are important
since we examine small fluctuations of bosonic fields from a stationary point.    
It is a known fact that if a Lie group is simply connected, its global structure is determined by the corresponding Lie algebra.
The global structure of the effective action/potential of a theory will be known by both its group theoretical nature 
and quantum field theory.
In a global aspect, $Spin(N)$ is the double-covering group of $SO(N)$, and thus, if $SO(N)$ gauge theory has a unique vacuum in its 
$SO(N)$ fundamental domain, the corresponding $Spin(N)$-gauge theory must have two exactly degenerate vacua,
though they cannot be distinguished by the Lie algebra ( namely, a locally defined quantity ) in general.

Our anomalous NG theorem has not only a locally characteristic aspect ( quasi-Heisenberg algebra, massive NG bosons, Weyl algebra )
but also some interesting global nature, which reflect some number theoretical aspects such as a fundamental group of covering group
or a Galois group.
Such a global aspect of our anomalous NG theorem directly reflects to symmetry between stationary points and geometry of stationary points,
given by the effective potential $V_{eff}$.
For our understanding of the global nature of an NG manifold/variety 
( probably some class of quotients of breaking schema, if it has a coset topology,
may have fixed points and singularities ) of the anomalous NG theorem, we need further investigation.
Some exotic mathematical nature of NG manifolds provides us a new subject of study on submanifolds in ( differential ) geometry.
The global character of an NG manifold, namely its compactness or the fundamental group, 
is understood by neglecting the local details of effective potential, 
and then convert our problem to a problem of Lie groups and homogeneous spaces.
Sometimes $G\to H$ gives a Riemannian/Hermitian symmetric space,
while it might be possible to generate a pseudo-Riemannian space as a result of symmetry breaking 
( this is not familiar from a context of physics ). 
Topological/global nature of pseudo-Riemannian spaces is still not yet understood enough in modern mathematics~[43]. 
Since there is a common understanding that a fundamental group is a Galois group~[72], 
our problem continues to the region of number theory.  
From this aspect of the global character of geometry of symmetry breaking, 
especially the Clifford-Klein form $\Gamma\backslash G/H$ is important since it has some nice properties,
and the case where $\Gamma$ is proper discontinuous and free is interesting for us.
If a breaking scheme $G\to H$ gives a Hermitian symmetric space $G/H$,
then it is a known fact that the $G/H$ has a uniform lattice $\Gamma$ ( i.e., $\Gamma\backslash G/H$ is compact ).

A more complicated situation will arise when we consider a breaking scheme 
under some generators of a Lie algebra are already broken by an explicit symmetry breaking parameter, 
namely so-called "explicit+spontaneous" symmetry breakings~[70].
( We give an example ( the kaon condensation model ) of it ( anomalous+explicit+spontaneous symmetry breaking ) later in this paper. )
In such situations, for example, a breaking scheme such as 
\begin{eqnarray}
SU(N) \to ({\bf Z}/N{\bf Z})^{\times} \simeq Gal({\bf Q}(\zeta_{N})/{\bf Q})
\end{eqnarray}
can take place. Here $\zeta_{N}$ is the $N$-th root of unity, and $({\bf Z}/N{\bf Z})^{\times}$ gives the center of $SU(N)$.
( Note that $\det[{\rm diag}(\underbrace{\zeta_{N},\cdots,\zeta_{N}}_{N})]=1$. )
Namely, it gives the following central extension:
\begin{eqnarray}
1 \to ({\bf Z}/N{\bf Z})^{\times} \to SU(N) \to PSU(N) \to 1.
\end{eqnarray}
( $PSU(N)$; projective special unitary group. )
$Gal({\bf Q}(\zeta_{N})/{\bf Q})$ is the Galois group of the cyclotomic extension.
This exact sequence is useful to consider a breaking scheme which gives a Grassmannian 
$SU(N)/SU(N-M)SU(M)$ and then successively ${\bf Z}/(N-M){\bf Z}\times {\bf Z}/N{\bf Z}$.
This breaking scheme may have a quite interesting mathematical implication in a quantum group
by utilizing the quasi-Heisenberg and quasi-Weyl algebra representations:
The discrete Heisenberg group ( all of the matrix elements of representation of a discrete Heisenberg group are integers ) 
is given by the algebraic relations of generators such that $xy=zyx$, $[x,z]=[y,z]=0$.
The author speculate this is the first time to find a quantum algebra in the NG theorem.
Needless to say, the relation $xy=zyx$ is consistent with the canonical Heisenberg algebra $[x,y]=1$
by setting $xy=z/(z-1)$ and $yx=1/(z-1)$ ( they recover the canonical commutation relation ).
Moreover, Bost and Connes construct a theory of dynamical system which gives a Galois group $Gal({\bf Q}(\zeta_{N})/{\bf Q})$ associated 
with a spontaneous symmetry breaking, and the partition function below the critical temperature is the Riemann zeta function~[7].
It is emphasized that their work has a strong connection with the Riemann hypothesis.
In fact, our theory of NG theorem contains some parts of algebraic aspects of their work naturally.
For example, a global nature of our anomalous NG theorem can give $Gal({\bf Q}(\zeta_{N})/{\bf Q})$.
Hence our theory of NG theorem might provide an approach toward the solution of the Riemann hypothesis:
This point will be discussed later in this paper.
We need a systematic investigation on the relation between several symmetry-breaking schema and Galois representations,
with a perspective of (non)commutative class field theory, i.e., the Langlands conjecture~[22,23,24,29,58,80]. 
From similar perspective, the following short exact sequences are also interesting:
\begin{eqnarray}
& & 1 \to {\bf Z}/2{\bf Z} \to {\rm Spin}(N) \to SO(N) \to 1,  \\
& & 1 \to {\bf Z}/2{\bf Z} \to {\rm Spin}^{\bf C}(N) \to SO(N)\otimes U(1) \to 1.
\end{eqnarray}
In fact, ${\rm Spin}(N)$ is a double covering group of $SO(N)$,
and ${\rm Spin}^{\bf C}(N)$ is its complexification.
Also, ${\rm Spin}(N,{\bf R})$ is the group in the theory of Clifford algebra.
${\bf Z}/2{\bf Z}$ is a Galois group.
The upper exact sequence is frequently used for an explanation of a Stiefel-Whitney class
which judges whether a manifold is orientable.
A central extension of Lie group by a discrete group corresponds to the covering space, 
directly related with the fundamental group.
Moreover, a central extension of Lie group induces a central extension of Lie algebra ( but its inverse is not true in general ).
The Lie's third theorem states that a simply connected Lie group exists for a given finite dimensional Lie algebra.
We make a brief comment on a central extension of Lie algebra~[63]. 
For example,
\begin{eqnarray}
0 \to {\bf R} \to \widetilde{{\rm Lie}(G)} \to {\rm Lie}(G) \to 0.
\end{eqnarray}
Here, $\widetilde{{\rm Lie}(G)}$ is the central extension of ${\rm Lie}(G)$ by ${\bf R}$.
Some literature given as our references discuss possible roles of central extensions to Lie brackets 
which might affect on the anomalous behavior of NG theorem.
It is a well-known fact that there is no nontrivial central extension if ${\rm Lie}(G)$ is semisimple.
A central extension may have a role when we consider a symmetry of a Kac-Moody group or a Heisenberg group. 
The following isomorphism is useful for us: $\pi_{1}(G/H)\simeq \pi_{1}(G)$ where $G$ is a connected Lie group,
and $H$ is a simply connected closed subgroup of $G$.
Thus, the nature of covering space which is implied by a central extension of a Lie group conserves under a breaking scheme $G\to H$,
and it is enough for us to consider a fundamental/Galois group of covering space of $G$.
Those covering groups and Galois groups describe symmetries of stationary points of NG sectors. 
For example, the set of stationary points inside the fundamental domain of $G/H$ acquires the symmetry of $\pi_{1}(G/H)$.

From our examination, there are functors of algebra cohomologies associated with a breaking scheme of anomalous NG theorem, 
such that,
\begin{eqnarray}
{\rm Lie \, algebra \, cohomology} \to {\rm Heisenberg \, algebra \, cohomology} \to {\rm Galois \, cohomology}.
\end{eqnarray} 
This is a remarkable fact since, for example, a Lie algebra cohomology describes the topological nature of underlying Lie group.
If a symmetry of a set of stationary points in an NG sector is a Galois type,
the set gives a Galois representation controlled by a Galois cohomolgy.
These cohomology, especially a Galois cohomology may have an overlap with an \'{e}tale cohomology 
since a Galois cohomology is a special case of \'{e}tale cohomology which implies an underlying algebraic variety.
This fact may help us to understand the underlying mechanism of the relations of those cohomologies and algebras.
The relationship between a Heisenberg algebra and a Galois group is a characteristic aspect of our anomalous NG theorem,
while other relations may be contained also in the normal NG theorem. 
A Galois group appears in various geometric examples but of particular interest here 
is geometric expressions of class field theory, several Galois representations, and \'{e}tale fundamental groups
by our anomalous NG theorem. 
It may be noteworthy to mention that the Abelianized part of the total Lie algebra reflects the flatness of the effective action/potential
of the theory, while the quasi-Heisenberg relation lifts partly the degeneracy of the vacua of the theory
along with some NG-bosonic coordinates. 
Since an apparent discrete symmetry between stationary points takes place in a massive NG-bosonic coordinate/space,
a Galois representation will be found in the space of a quasi-Heisenberg relation.

Now, we list some breaking schema interesting for us from the context of this paper: 
\begin{eqnarray}
& & SU(4) \to SU(2)\otimes SU(2) \simeq SO(4) \to SU(2)_{diag} \to U(1),  \\ 
& & SU(5) \to SU(3)\otimes SU(2)\otimes U(1) \to U(1)\otimes U(1)\otimes U(1)\otimes U(1),  \\
& & SU(6) \to SU(3)_{L}\otimes SU(3)_{R} \to SU(3)_{V} \to U(1)\otimes U(1),  \\
& & SO(10) \to SU(4)\otimes SU(2)\otimes SU(2),  \\
& & E_{6} \to SO(10)\otimes U(1) \to SU(5), \\
& & E_{8}-{\rm ferromagnet}, \quad ( {\rm experimentally \, observed \, spin \, system} ), \\
& & G_{2} \to SO(4),  \\
& & SU(N) \to SU(N-M), \quad ( {\rm Stiefel \, manifold} ), \\
& & SO(N) \to SO(N-M), \quad ( {\rm Stiefel \, manifold} ), \\
& & SU(N) \to SU(N-M)\otimes SU(M), \quad ( {\rm symmetric \, space, \, Grassmann} ), \\
& & SO(N) \to SO(N-M)\otimes SO(M), \quad ( {\rm symmetric \, space, \, Grassmann} ),  \\
& & Spin(6) = SU(4) \to something,  \\
& & Spin(4,2) = SU(2,2) \to something.
\end{eqnarray}
In those examples, 
if a symmetry breaking takes place under a breaking scheme in which some elements of the Cartan subgroup of the total group remains unbroken,
then it is trivial that a ( quasi ) Heisenberg algebra arises.
For example, the breaking scheme $SU(N) \to SU(N-M)\otimes SU(M)$ will take place by an order parameter of the form
diag$(\underbrace{a,\cdots,a}_{N-M},\underbrace{b,\cdots,b}_{M})$ which should be proportional to a linear combination of VEVs of the Cartan subalgebra of $SU(N)$.
A large part of breaking schema listed above are fall into this class of symmetry breakings.
The breaking scheme $SU(N)_{L}\otimes SU(N)_{R}\to SU(N)_{V}$ is famous in a chiral symmetry breaking of left-right symmetric $N$-flavor model.
In this case, the Lie algebra one considers is
\begin{eqnarray}
& & [\theta^{a}T^{a}\otimes 1 + \varphi^{b}T^{b}\otimes\sigma^{3}, \Phi], \nonumber \\ 
& & \Phi\propto \sigma^{1}, \quad  \theta^{a}T^{a}\otimes 1 \in {\rm Lie}(SU(N)_{V}), \nonumber \\
& & (\theta^{a}+\varphi^{a})T^{a}\otimes\frac{1+\sigma^{3}}{2} \in {\rm Lie}(SU(N)_{L}), \nonumber \\
& & (\theta^{a}-\varphi^{a})T^{a}\otimes\frac{1-\sigma^{3}}{2} \in {\rm Lie}(SU(N)_{R}).
\end{eqnarray}
Here, $\theta^{a}T^{a}\otimes 1$ remains symmetric while $\varphi^{b}T^{b}\otimes\sigma^{3}$ is broken.
Hence, the VEV takes its value toward $\sigma^{1}$ direction ( as you know, $\gamma^{0}$ is frequently used, while $\sigma^{3}\to \gamma_{5}$ ),
and this case does not belong to the class of diagonal breaking we have studied in this paper.

$E_{8}$ might have several exotic breaking schema due to its Dynkin diagram and Cartan martrix
while our observation of anomalous NG theorem in a generic case should valid to it.
An example of spin system of $E_{8}$ symmetry has been observed experimentally quite recently~[6,9,88].
Zamolodchikov seems to use his theory ( affine Toda field theory of $E_{8}$ ) to give a mass spectrum of mesons, two quark ( kink ) bound states
and thus his theory is constructed in a (1+1)-dimensional model, 
though some part of the mechanism of generating an $E_{8}$ spectrum
is independent from the dimensionality of a system, determined by the Lie algebra Lie$(E_{8})$. 
Hence an $E_{8}$ spin system has an importance from its own right, beyond its dimensionality.
( A breaking scheme of $E_{8}$ is also interesting for us from the context of the Kazhdan-Lusztig-Vogan polynomials for $E_{8}$~[38,39,51]. )

\subsection{Riemannian and Hermitian Symmetric Spaces}

If a breaking scheme $G\to H$ gives a symmetric space~[32], 
several geometric properties will be introduces to our NG theorem more concretely.
Especially, a local nature ( Lie algebra ) and a global structure ( Lie group ) is bridged more clearly.
First, we summarize the basic well-known fact of a symmetric space.
Let $M$ be a symmetric space. 
A Lie group $G$ acts transitively on $M$.
In addition, an involution $s$ is defined for any local point of $M$, as an automorphism of $M$,
and $s$ acts on any group element of $G$ as an adjoint $sgs^{-1}$.
This involution is an automorphism of $G$ itself, and of course it acts on ${\rm Lie}(G)$ as an automorphism.
Then ${\bf g}={\rm Lie}(G)$ is decomposed into ${\bf h}+{\bf m}$ by their eigenvalues of operations of $s$ 
( ${\bf h}\to +1$, the Cartan subalgebra, and ${\bf m}\to -1$ ).
Hence, by the number of odd elements ${\bf m}$, the relations $[{\bf h},{\bf h}]\subset {\bf h}$, 
$[{\bf h},{\bf m}]\subset{\bf m}$, $[{\bf m},{\bf m}]\subset{\bf h}$ are immediately obtained
( this is a kind of grading of the algebra by the set of odd elements ${\bf m}$ ).
${\bf m}$, an orthogonal complement space of ${\bf h}$, is isomorphic with $TM$,
and a curve $t\to e^{it{\bf m}}\cdot o$ ( $o$: a point of $M$, $t\in{\bf R}$ ) is geodesic.
As stated above, the tangent bundle of a Riemannian manifold has $O(n)$ as the structure group.
If a spontaneous symmetry breaking $G\to K$ gives a symmetric space, $M=G/K=e^{i{\bf m}}$, and the Cartan subalgebra remains unbroken,  
then the NG manifold is expanded only by the basis of broken generators ${\bf m}$ 
which is isomorphic with $TM=T(G/K)$ which may have the structure group $O(n)$, 
and the NG bosons $\{\chi^{A}\}$ as the local coordinate system expressed by $\Phi \to e^{i\chi^{A}m^{A}}\Phi$
are all geodesic, whether the normal or anomalous cases of NG theorem.
The VEVs $\langle{\bf h}\rangle\ne 0$ are always normal ( vertical ) with the tangent space given by the NG boson space ${\bf m}$.
A quasi-Heisenberg algebra is globally defined, inside the linear space $TM$.
Since an NG boson gives a geodesic over the manifold $G$, a Jacobi field is associated along with the geodesic curve. 
The curvature tensor is given by $R(X,Y)Z=[[X,Y],Z]\subset{\bf m}$ 
( here, $X,Y,Z\in{\bf m}$ and we have used the physics convention of $g=e^{i{\bf m}}$, where $i=\sqrt{-1}$ ) at the origin.
If an order parameter $\Phi$ belongs to ${\bf h}$, 
then the Riemann curvature $R$ appears at the fourth order displacement of a Lagrangian or an effective potential 
caused by broken generators.
Here, a symmetric space is defined by a Lie algebra, thus it does not depend on details of the manifold. 
For example, $SU(N) \to SU(N-M)\otimes SU(M)$ is a symmetric space, thus broken generators form the algebra $[X^{a},X^{b}]\subset S^{c}$,
and if $\langle S^{c}\rangle\ne 0$, a ( quasi ) Heisenberg algebra arises.
In this case, Lie$SU(N)$ is projected to the quasi-Heisenberg algebra globally via our anomalous NG theorem.
Especially interesting for us is the fact that a unitary group $U(N)$ is a Hermitian manifold
( keeps a Hermitian structure of a quadratic form ), 
while a Heisenberg group is possibly be described by a Riemannian manifold.
Hence, our anomalous NG theorem may bridge between a complex ( K\"{a}hler ) structure and 
a quantized symplectic ( quasi-Heisenberg ) structure in a symmetric space.

\subsection{A Heisenberg Group as the Symmetry of the Beginning}

If a Heisenberg group ( sometimes used in a flavor dynamics, a flavor-symmetry breaking ) 
is an internal symmetry from the beginning of a theory, and if one considers its spontaneous symmetry breaking,
a situation similar with the cases of semisimple classical Lie groups takes place,
since a set of VEVs of a Heisenberg algebra can again give a Heisenberg algebra.
For example, 
\begin{equation}
\Xi = \left(
\begin{array}{ccc}
1 & 0 & 0 \\
0 & 0 & 0 \\
0 & 0 & 0 
\end{array}
\right) \quad {\rm or} \quad \left(
\begin{array}{ccc}
0 & 1 & 0 \\
0 & 0 & 0 \\
0 & 0 & 0 
\end{array} 
\right) \quad {\rm or} \quad \left(
\begin{array}{ccc}
0 & 0 & 1 \\
0 & 0 & 0 \\
0 & 0 & 0 
\end{array}
\right),
\end{equation}
with an action of a Heisenberg group $G$ ( see, (35) ) from the left side gives $G\Xi=\Xi$, namely $\Xi$ is $G$-singlet,
can be utilized to make an invariant theory. 
This form of $\Xi$ may be attractive for an attempt to generate a flavor degree of freedom:
For example, the following $\widetilde{\Xi}$ which breaks a Heisenberg-group symmetry can generate a flavor hierarchy by an action of $G$:
\begin{equation}
\widetilde{\Xi} = 
\left(
\begin{array}{ccc}
1 & 0 & 0 \\
0 & \epsilon & 0 \\
0 & 0 & \epsilon^{2} 
\end{array}
\right), \quad |\epsilon| \ll 1.
\end{equation} 
Therefore, one can consider a symmetry breaking which generates a flavor hierarchy via our anomalous NG theorem:
\begin{flushleft}
SU(2) ( SU(N) ) $\to$ Heisenberg group $\to$ something. 
\end{flushleft}

An important issue is coming from the fact that the Killing form of a nilpotent Lie algebra is identically zero,
while the Killing form of a solvable Lie group gives a different result from it.
A quotient of Heisenberg group gives a solvmanifold~[3].
A fermion model of solvable Lie group symmetry with a dynamical symmetry breaking which generates a sigma model might be possible. 
In a bosonic field theory of a compact Lie group, the Killing form is negative definite,
while in the case of indefinite signature of Killing form, a model Lagrangian should be regarded as an "analytic continuation" from
a physical model.

\subsection{Kac-Moody Algebras, Generalized Kac-Moody Algebras, and Affine Lie Algebras}

It should be noticed that a Lie group which derives a N\"{o}ther charge can be treated ( at least, formally in a certain sense ) 
as an internal symmetry of a model Lagrangian. 
We assume a Kac-Moody Lie algebra~[25,37] has its corresponding Lie group via an exponential mapping, surjectively,
even though this assumption is sometimes violated,
and it is a non-trivial issue to define a Haar measure for a Kac-Moody group.
This issue affects to define a path-integral measure over a Kac-Moody group.
From our perspective of this paper, the most interesting fact is that, after taking VEVs of brackets of a Kac-Moody algebra
( now we have an infinite number of conserved charges ),
especially an affine Lie algebra, we obtain a direct sum of quasi-Heisenberg algebras of polynomial growths
( would be called as an infinite-dimensional quasi-Heisenberg algebra ) and Abelianized subalgebras.
The method to construct a Kac-Moody algebra 
( a Cartan subalgebra, a root system, a Cartan matrix, etc. )
is parallel with the general theory of finite-dimensional simple Lie algebras~[37]:
Hence, our discussion given above can straightforwardly be applied to several cases of Kac-Moody algebras.
Our theory can be regarded as a higher-dimensional version of a ( affine ) Toda field theory,
in which a scalar field of the theory takes its value on the Cartan subalgebra of Kac-Moody algebra.
After a diagonal breaking scheme takes place,
such a higher-dimensional version of affine Toda field theory-like model acquires a quasi-Heisenberg algebra.
A case beyond a diagonal breaking causes more complicated result.
An affine Lie algebra can be interpreted as a special form of trivial fiber bundle, ${\bf g}\otimes {\bf C}[t,t^{-1}]$.
Hence a $G$-bundle and a Maurer-Cartan form of Cartan geometry can, at least formally, be considered.

Another different perspective is coming from an ${\cal N}=2$ superconformal algebra, especially the so-called coset construction.
It is constructed by the affine Kac-Moody algebra of $SU(2)$ at level $l$,
\begin{eqnarray}
& & [h_{m},h_{n}] = 2ml\delta_{n+m,0}, \quad [e_{m},f_{n}] = h_{m+n} + ml\delta_{m+n,0},  \nonumber \\
& & [h_{m},e_{n}] = 2e_{m+n}, \quad [H_{m},f_{n}] = -2f_{m+n},
\end{eqnarray}
with an associated set of complex Grassmann variables.
Thus, if a generic Lagrangian of an NG sector ( for example, see (104) ) is pairwisely decomposed via taking VEVs, 
as the direct sum of VEVs of algebra of Lie$(SL(2))$-triple $(h, e, f)$,
then the algebra inside the Lagrangian is a finite and special version of the $SU(2)$ affine Lie algebra:
The algebra arised in the generic NG-boson Lagrangian can be embedded into the superconformal algebra.
This type of discussion is useful for us to consider several relations between our generic NG-boson Lagrangian and other theoretical models. 
It is a known fact that a class of infinite-dimensional simple linearly compact Lie superalgebras contain the Standard Model gauge group
$SU(3)\otimes SU(2)\otimes U(1)$ as the algebra of level zero.
Our argument presented here has some similarity with such a situation.
The important issue is to know how geometries of these Lie groups are related with each other.
It might be possible to obtain affine Lie groups starting from Lie algebras of Riemann/Hermitian symmetric spaces.

\subsection{Graded Lie Algebras and Lie Superalgebras}

Lie superalgebras and Lie supergroups have quite interesting characters, and they have importances in their own right~[36,89],
while they also acquire attention from some particle phenomenological point of view.
An extension of our anomalous NG theorem to supersymmetric theory is an interesting subject for us to complete our theorem.
This will be done in another paper by the author ( in preparation ), 
and here we will see some perspectives especially from mathematics.
For example, we can consider the following diagram:
\begin{flushleft}
Lie superalgebra $\to$ Heisenberg superalgebra ( bosonic/fermionic ) $\to$ Galois supergroup. 
\end{flushleft}
The notion of Galois supergroup is not strange, if we consider a non-trivial central extension to give the supergroup.
Via an effective potential and an order parameter, notions of supermodules, superschemes would be introduced in our theory of NG theorem.
For our context of this paper, the following diagram is considered:
\begin{flushleft}
Lie superalgebras $\to$ Lie supergroups $\to$ supergroup-schemes $\to$ superschemes $\to$ super-\'{e}tale cohomology 
$\to$ super-Galois representations,
\end{flushleft}
and,
\begin{flushleft}
super-sheaves and supermodules $\to$ superschemes $\to$ perverse super-sheaves $\to$ super-intersection cohomology $\to$ stratified super-Morse theory,
\end{flushleft}
and,
\begin{flushleft}
supergroups $\to$ Maurer-Cartan superforms $\to$ Cartan supergeometry
\end{flushleft}

\section{The Effective Potential Formalism}

Let us examine the effective action $\Gamma_{eff}$ and effective potential $V_{eff}$ of a general situation.
( See the book of Kugo~[49]. )
Let $\Phi$ be a matrix order parameter.
$\Phi$ can be regarded as a left $G$-module~[20,30]: 
$g(\Phi_{1}+\Phi_{2})=g\Phi_{1}+g\Phi_{2}$, $g\in G$, $\Phi_{1},\Phi_{2}\in \Phi$.
$\Phi$ is assumed as $G$-equivariant.
Both $G$ and $\Phi$ are defined over the same field $F$, usually ${\bf C}$ or ${\bf R}$, 
and thus $G$ and $\Phi$ acquire the same topology with $F$.
Since $G$ acts on $\Phi$ continuously, $\Phi$ is a topological $G$-module.
A group (co)homology $H_{n}(G,\Phi)$ and $H^{n}(G,\Phi)$ can be considered 
by modules generated by $g\Phi_{1}-\Phi_{1}$ ( difference ), $\Phi_{1}\in\Phi$ and $g\in G$, under the systematic manner. 
There are several nontrivial issues in our situation due to the nature of quantum field theory
( $\Phi$ is a quantum field, not exactly the same with ${\bf R}$, ${\bf C}$, or ${\bf Z}$ ), as a physical system.
If we regard the effective potential $V_{eff}$ as a scheme, or when $V_{eff}$ defines an algebraic variety,
then the (co)homology groups $H_{n}(X,{\cal O}_{X})$ and $H^{n}(X,{\cal O}_{X})$ ( ${\cal O}_{X}$; a sheaf ) can also be considered~[30,31].
( Such a cohomology group is introduced anywhere we meet a sheaf in our theory. 
The \'{e}tale cohomology is a cohomology theory of sheaves in the \'{e}tale topology~[53].
It may be possible to apply the method of \'{e}tale cohomology to study a topological nature of our NG theorem. )
If $V_{eff}$ is included in a line bundle, the Borel-Weil theory can be applied~[43].

Let us consider a linear displacement of a field $\Phi$ caused by a conserved charge $Q^{A}$:
\begin{eqnarray}
\delta_{A}\Phi &=& [Q^{A},\Phi] = \theta^{A}T^{A}\Phi, \quad ( A = 1, \cdots, N ), \quad T^{A}\in {\rm Lie}(G).
\end{eqnarray}
A typical example is the chiral $\gamma_{5}$ transformation: $[Q^{5},\bar{\psi}i\gamma_{5}\psi]=-2\bar{\psi}\psi$,
and $\langle \bar{\psi}\psi\rangle\ne 0$ gives an order parameter with a fixed phase of the chiral rotation.
By takings its VEV, we get
\begin{eqnarray}
\langle 0|[Q^{A},\Phi]|0\rangle &=& \langle 0|\theta^{A}T^{A}\Phi|0\rangle.
\end{eqnarray}
With taking into account $Q^{A}|0\rangle =0$ ( symmetric ) and $Q^{A}|0\rangle\ne 0$ ( broken ),
usually we conclude that $\langle 0|\theta^{A}T^{A}\Phi|0\rangle=0$ ( symmetric ) and $\langle 0|\theta^{A}T^{A}\Phi|0\rangle\ne 0$ ( broken ).
However, in the case of $SU(2)\to U(1)$ of a ferromagnet, 
$\langle 0|Q^{z}|0\rangle = \int d^{3}{\bf x}\langle 0|j^{z}_{0}(x_{0},{\bf x})|0\rangle \ne 0$ may take place even though $Q^{z}$ is unbroken. 
Next, we give a formal expansion of $V_{eff}[\Phi]$ by the set of vectors of ${\rm Lie}(G)$ around the VEV:
\begin{eqnarray}
V_{eff}[\Phi] &=& V_{eff}[v] 
+ \Bigl(\frac{\partial V_{eff}}{\partial\Phi}\Bigr)_{\Phi=v}(\theta^{A}T^{A}\Phi) 
+ \frac{1}{2!}\Bigl(\frac{\partial^{2} V_{eff}}{\partial\Phi^{2}}\Bigr)_{\Phi=v}(\theta^{A}T^{A}\Phi)^{2} + \cdots.
\end{eqnarray}  
Here $v$ implies a certain type of VEV of $\Phi$.
The effective potential $V_{eff}$ belongs to a germ of a sheaf of smooth function ${\cal O}_{D}$ ( $D$: a domain ). 
Thus, after taking a VEV of this expansion, the stationary condition gives the following criterion
( Eqs. (16)-(18) of the paper of Goldstone, Salam, and Weinberg~[27] ):
\begin{eqnarray}
\Bigl(\frac{\partial^{2} V_{eff}}{\partial\Phi^{2}}\Bigr)_{\Phi=v}\langle 0|[Q^{A},\Phi]|0\rangle =
\Bigl(\frac{\partial^{2} V_{eff}}{\partial\Phi^{2}}\Bigr)_{\Phi=v}\langle 0|\theta^{A}T^{A}\Phi|0\rangle =0.
\end{eqnarray}
This equation is coming from the N\"{o}ther theorem of a conserved current combining with the stationary a condition of $V_{eff}$. 
Since the zeroth cohomology group of a Lie algebra is defined by the set of invariants ( annihilated ) under the algebra operation on a module~[20],
\begin{eqnarray}
H^{0}({\rm Lie}(G),M) &=& M^{{\rm Lie}(G)} = \bigl\{ m\in M| gm=0, \, \forall g\in {\rm Lie}(G) \bigr\}.
\end{eqnarray}
Thus the matrix $\frac{\partial^{2}V_{eff}}{\partial\Phi^{2}}$ is interpreted as an "effective" invariant module, 
the zeroth cohomology of the Lie algebra, via taking VEVs in the quantum theory.
$\frac{\partial^{2}V_{eff}}{\partial\Phi^{2}}$ is regarded to take its value in the sheaf of germs of continuous functions.
( For example, when $M={\bf R}$, any connected compact semisimple Lie group $G$ has
$H^{0}({\rm Lie}(G),{\bf R})={\bf R}$, $H^{1}({\rm Lie}(G),{\bf R})=H^{2}({\rm Lie}(G),{\bf R})=0$. )
In the usual case, the equation (68) has exactly the same dimension with the number of broken generators and closed in the linear space,
while in the case such as a ferromagnet, $\langle 0|Q^{z}|0\rangle\ne 0$.
Note that this VEV will be rewritten after a change of basis set of Lie algebra, an algebra homomorphism, 
i.e., $\langle 0|Q^{z}|0\rangle=\langle 0|Q^{x'}+Q^{y'}+Q^{z'}|0\rangle \ne 0$,
and thus the final result obtained from any calculation must not depend on the choice of Lie algebra representation, 
$\langle Q^{z}\rangle\ne 0$. 
This fact means that a rotation of the frame $(x,y,z)$ must not affect on the physical content of this equation, of course.
The important point is that this equation (68) cannot be written down only by the set of broken generators in the ferromagnetic case: 
This case apparently breaks the condition of proof of the ordinary NG theorem,
and one cannot conclude the existence of a zero-mass bosonic particle.
In other words, the NG boson subspace interacts with the "symmetric" subspace in a breaking scheme caused by a quantum effect:
The author argues that this is a kind of quantum geometry.
Thus, the mass matrix of NG bosons,
\begin{eqnarray}
\Bigl(\frac{\partial^{2} V_{eff}}{\partial\Phi^{2}}\Bigr)_{\Phi=v} &=& \Delta^{-1}_{F}(p=0)
\end{eqnarray} 
( $\Delta_{F}(p)$; a matrix Feynman propagator of NG bosons ) "should" have a nonzero value. 
This is just the mechanism of famous Nielsen-Chadha anomaly in the NG theorem~[1,8,34,55,77,81,82,83,84].
Simultaneously, it is also clear from our discussion, a Lorentz-invariant system with a breaking scheme $G/H$ which gives a symmetric space
never has a "spontaneous violation" of the ordinary/normal NG theorem.
( The Nielsen-Chadha anomaly never takes place. )
Hence, the ordinary NG theorem is protected by the Coleman-Mandula theorem of $S$-matrix.
Not only the mass matrix of NG bosons, but the dispersion relations themselves should be derived from $\frac{\partial^{2}V_{eff}}{\partial\Phi^{2}}$.
It should be mentioned that there might be a similar situation in a relativistic model
with a Lorentz-violating parameter:
For example, the following VEV in an $SU(2)$ isospin space could be considered, 
\begin{eqnarray}
\langle \bar{\psi}\tau_{3}\psi\rangle \ne 0,
\end{eqnarray}
in a NJL type model. In this VEV, $\tau_{3}$ is symmetric while $(\tau_{1},\tau_{2})$ is broken. 
It might give a similar situation with the ferromagnet with $\langle Q^{3}_{isospin}\rangle\ne 0$.

Let us discuss further on $V_{eff}$.
From the general theory of effective action, the displacement (65) derives the following equation, 
from a second-order derivative of the effective potential $V_{eff}$ after taking a VEV:
\begin{eqnarray}
0 &=& \langle (V''_{eff})_{1}\theta^{1}T^{1}\Phi + \cdots + (V''_{eff})_{N}\theta^{N}T^{N}\Phi \rangle \nonumber \\
&=& \langle (V''_{eff})_{1}\theta^{1}T^{1}v + \cdots + (V''_{eff})_{N}\theta^{N}T^{N}v \rangle,  \\
V''_{eff} &=& \frac{\delta^{2} V_{eff}[\Phi]}{\delta\Phi^{2}}, \\
\Phi &=& c_{1}T^{1} + \cdots + c_{N}T^{N}, \quad \{c_{j}\}\in{\bf C}, \, ( j= 1, \cdots, N ).
\end{eqnarray} 
Here $v\in M_{n}({\bf C})$ ( matrix ) indicates a VEV. 
$\Phi$ is a $G$-module expanded by the basis of ${\rm Lie}(G)$.
Some normalization condition for $\Phi$ is set aside for a while.
Since $T^{A}\Phi$ is cased by taking adjoints, the above equation implicitly contains the root space of the Lie algebra
( i.e., from $[{\bf h},{\bf g}]=\lambda({\bf h}){\bf g}$, $\lambda({\bf h})$; root, ${\bf h}$; Cartan subalgebra, ${\bf g}\in {\rm Lie}(G)$ ),
and the corresponding Weyl group acts implicitly.
In a case of Riemannian symmetric space, the adjoints, the Killing form, and the Jacobi field are obtained from its root system
( see the book of Helgason~[32] ).
Then they are related with a harmonic mapping and a harmonic analysis.
Later, we will mention that an NG boson gives a geodesic in a case of Riemannian symmetric space.
Note that the ${\rm Lie}(G)$ itself can be regarded as a $G$-module, by satisfying the axiom of a $G$-module 
with a certain type of group operations $G\times {\rm Lie}(G)\to{\rm Lie}(G)$.
This formula (72) includes the "off-diagonal" contributions of the second-order derivative of $V_{eff}$
in the space of Lie algebra generators
which may cause mode-mode couplings of NG bosons:
Later, we will observe that a mode-mode coupling between bosonic fields modifies dispersion relations of NG bosons
and as a consequence, an NG boson acquires a finite mass. 
Needless to say, the bases $T^{A}\in {\rm Lie}(G)$ are always linearly independent.
For example, in a case of $SU(2)\to U(1)$,
\begin{eqnarray}
0 &=& (V''_{eff})_{1}\langle[Q^{1},\Phi]\rangle + (V''_{eff})_{2}\langle[Q^{2},\Phi]\rangle + (V''_{eff})_{3}\langle[Q^{3},\Phi]\rangle,  \\
\Phi &=& c_{1}\sigma^{1} + c_{2}\sigma^{2} + c_{3}\sigma^{3}.
\end{eqnarray}
Here, $\Phi\in{\bf 2}$-representation.
We write the Lie brackets in the above equation as follows:
\begin{eqnarray}
& & \langle [Q^{1},c_{2}\sigma^{2}]\rangle = (c_{2})_{1}\sigma_{3}, \quad 
\langle[Q^{1},c_{3}\sigma^{3}]\rangle  = (c_{3})_{1}\sigma_{2}, \nonumber \\
& & \langle[Q^{2},c_{1}\sigma^{1}]\rangle  = (c_{1})_{2}\sigma_{3}, \quad
\langle[Q^{2},c_{3}\sigma^{3}]\rangle  = (c_{3})_{2}\sigma_{1}, \nonumber \\
& & \langle[Q^{3},c_{1}\sigma^{1}]\rangle  = (c_{1})_{3}\sigma_{2}, \quad
\langle[Q^{3},c_{2}\sigma^{2}]\rangle  = (c_{2})_{3}\sigma_{1}.
\end{eqnarray}
Then we get
\begin{eqnarray}
0 &=& \bigl\{ (V''_{eff})_{1}(c_{2})_{1} + (V''_{eff})_{2}(c_{1})_{2}\bigr\}\sigma_{3} \nonumber \\
& & + \bigl\{ (V''_{eff})_{2}(c_{3})_{2} + (V''_{eff})_{3}(c_{2})_{3}\bigr\}\sigma_{1}  \nonumber \\
& & + \bigl\{ (V''_{eff})_{1}(c_{3})_{1} + (V''_{eff})_{3}(c_{1})_{3}\bigr\}\sigma_{2} 
\end{eqnarray}
Due to the linear independence of $\sigma_{1,2,3}$, all of the coefficients vanish.
If $\langle[Q^{1},\Phi]\rangle\ne 0$ and/or $\langle[Q^{2},\Phi]\rangle\ne 0$,
both of them have contributions to rotate $\Phi$ to the third direction proportional to $\sigma^{3}$, 
and there is a freedom to take $(V''_{eff})_{1,2,3}$ finite in those vanishing coefficients
( the case of our anomalous NG theorem of a ferromagnet ),
while if $\langle[Q^{1},\Phi]\rangle = \langle[Q^{2},\Phi]\rangle = 0$, then $(V''_{eff})_{1,2,3}$ will vanish independently 
( the case of ordinary NG theorem ).
More general case is understood by the algebra we have discussed in the previous section:
At least in the case of quasi-Heisenberg algebra,
a mode-mode coupling takes place in $V_{eff}$ which becomes apparent from its second-order derivatives.
Hence, we arrive at the following theorem:
\begin{flushleft}
{\bf Theorem}:
{\it Any type of mode-mode coupling between NG bosons modifies their dispersion relations and mass spectra,
gives an anomalous behavior of the NG theorem.}
\end{flushleft}
We also yield another simple but important result:
\begin{flushleft}
{\bf Theorem}:
{\it Any spontaneous symmetry breaking of isolated $U(1)$ Abelian Lie group cannot give an anomalous behavior of the NG theorem,
due to the lack of mode-mode coupling.}
\end{flushleft}

In the theory of itinerant (anti)ferromagnetism of Moriya~[56,57], 
he pointed out that a mode-mode coupling of magnetic ( i.e., spin ) fluctuations
is important which can be experimentally observed by the method of magnetic resonance. 
Our anomalous NG theorem of a ferromagnet may have a physical implication in the Moriya theory.
If the mass of massive NG boson is controlled as a function of a strength external field or temperature,
the energy split of two NG bosons might be made small relative to the characteristic energy scale.
In the vicinity of the critical region, those bosons acquire special importance. 
In the theory of Moriya, the method of self-consistent renormalization theory is employed,
which takes into account diagrammatically higher-order interactions between electrons. 
It is interesting from our context that how such higher-order interactions affect the mass spectra of NG bosons, 
simultaneously to the value of $T_{c}$ and correlation lengths.

In a composite particle model, a Schwinger-Dyson equation determines an order parameter which is non-local,
for example a two-point function $\langle T\phi(x)\phi(y)\rangle\ne 0$~[15].
In that case, the gap equation is derived by ( in a translation-invariant case )
\begin{eqnarray}
\frac{\delta V_{eff}(G(p))}{\delta G(p)} &=& 0.
\end{eqnarray}
Here, $G(p)$ is a propagator.
Now, the mass matrix of NG bosons may be determined by the examination of the following equation:
\begin{eqnarray}
\sum^{N}_{A=1}\Bigl(\frac{\delta^{2}V_{eff}(G(p))}{\delta G(p)^{2}}\Bigr)_{A}T^{A}G(p) &=& 0.
\end{eqnarray}
Here, $T^{A}G(p)$ imply VEVs, for example, $\langle\bar{\psi}T^{A}\psi\rangle$. 
Now, the Lie group $G$ acts on the propagator $G(p)$ nontrivially,
though the algebra we consider here is an adjoint type, and thus this equation is essentially the same with (68) as an equation of Lie algebra.
Hence we obtain a mathematically similar equation with (72).
We conclude that the essential part of the basis of our anomalous NG theorem is the same in case of both composite and elementary fields.

\section{The Kaon Condensation Model}

Let us start from the following $SU(2)$ Higgs-Kibble-type model Lagrangian:
\begin{eqnarray}
{\cal L} &=& -\frac{1}{4}{\rm tr}F^{a}_{\mu\nu}F^{a\mu\nu} + \tilde{D}^{\dagger}_{\nu}\Phi^{\dagger}\tilde{D}^{\nu}\Phi 
+ m^{2}_{0}\Phi^{\dagger}\Phi - \frac{\lambda}{2}(\Phi^{\dagger}\Phi)^{2} + \epsilon m^{2}_{ex}\Phi^{\dagger}\sigma^{3}\Phi,  \nonumber \\
\tilde{D}_{\nu} &=& \partial_{\nu}-i\mu\delta_{0\nu}-\frac{i}{2}g\sigma^{a}A^{a}_{\nu},  
\end{eqnarray}
Here, we introduce $SU(2)$-gauge fields $A^{a}_{\nu}$ ( $a=1,2,3$ ), 
and the mass term $\epsilon m^{2}_{ex}\Phi^{\dagger}\sigma^{3}\Phi$ with  
a relatively small parameter $\epsilon$ explicitly breaks the symmetry. 
Since this explicit symmetry breaking parameter breaks the symmetries of $\sigma^{1}$ and $\sigma^{2}$,
both of them acquire finite masses.
The complex bosonic field is defined as $\Phi\equiv (\phi_{1},\phi_{2})^{T}$ as a ${\bf 2}$-representation, 
and $\mu$ is a Lorentz-symmetry violating chemical potential. 
Let us mention the fact that the anomalous behavior of NG bosons cannot be understandable by the hypercharge model of $U(2)$
used in the paper of Schaefer et al~[77]:
Due to the special nature of $SU(2)$, this model has an additional symmetry, so-called "custodial symmetry" of $SU(2)$.
Hence the breaking scheme of this model is $SU(2)\otimes SU(2)\simeq SO(4)\to SU(2)_{diag}$.
This breaking scheme is essentially the same with the Lie algebra of $SU(N)_{L}\otimes SU(N)_{R}\to SU(N)_{V}$ discussed at (61).
Namely, $[\theta^{a}\sigma^{a}\otimes 1+\varphi^{b}\sigma^{b}\otimes\sigma^{3}, \sum^{3}_{j=1}a_{j}\sigma^{j}]$ will be examined.
We consider the broken and symmetric generators carefully according to this breaking scheme.
First, let us consider the case where all of the gauge fields are dropped from this model.
We assume the model chooses $\Phi_{0}=\langle\Phi\rangle=(0,v)^{T}/\sqrt{2}$ as one of its vacua:
Then we yield the following relation for the VEV $v$:
\begin{eqnarray}
0 &=& \frac{\partial V^{(tree)}_{eff}}{\partial v}, \\
V^{(tree)}_{eff} &=& -\frac{1}{2}(\mu^{2}+m^{2}_{0}-\epsilon m^{2}_{ex})v^{2} + \frac{\lambda}{8}v^{4},  \\
v &=& \pm\sqrt{\frac{2(\mu^{2}+m^{2}_{0}-\epsilon m^{2}_{ex})}{\lambda}}.
\end{eqnarray}
Now a small displacement around the vacuum solution we have obtained, 
consists with the amplitude mode and three NG modes, is described by the following 't Hooft parametrization:
\begin{equation}
\Phi = \frac{1}{\sqrt{2}}
\left(
\begin{array}{c}
\chi_{2} + i\chi_{1} \\
v+\psi -i\chi_{3} 
\end{array}
\right)
\end{equation}
Then we get
\begin{eqnarray}
{\cal L} &=& \frac{1}{2}\Bigg[ 
\tilde{\partial}^{\dagger}_{\nu}\psi\tilde{\partial}_{\nu}\psi +
\tilde{\partial}^{\dagger}_{\nu}\chi_{a}\tilde{\partial}_{\nu}\chi_{a} + \mu^{2}v^{2}  \nonumber \\
& & \quad 
+ i\Bigl( \tilde{\partial}^{\dagger}_{\nu}\chi_{2}\tilde{\partial}_{\nu}\chi_{1}
- \tilde{\partial}^{\dagger}_{\nu}\chi_{1}\tilde{\partial}_{\nu}\chi_{2} \Bigr)    
+ i\Bigl( \tilde{\partial}^{\dagger}_{\nu}\chi_{3}\tilde{\partial}_{\nu}v
- \tilde{\partial}^{\dagger}_{\nu}v\tilde{\partial}_{\nu}\chi_{3} \Bigr)   \nonumber  \\
& & \quad - i\Bigl( \tilde{\partial}^{\dagger}_{\nu}\psi\tilde{\partial}_{\nu}\chi_{3}
- \tilde{\partial}^{\dagger}_{\nu}\chi_{3}\tilde{\partial}_{\nu}\psi \Bigr)  
+ \tilde{\partial}^{\dagger}_{\nu}\psi\tilde{\partial}_{\nu}v
+ \tilde{\partial}^{\dagger}_{\nu}v\tilde{\partial}_{\nu}\psi \Bigg]    \nonumber \\
& & \quad 
+ \frac{m^{2}_{0}}{2}\bigl( \chi^{2}_{1} + \chi^{2}_{2} + \chi^{2}_{3} + (v+\psi)^{2} \bigr)    \nonumber \\
& & \quad 
- \frac{\lambda}{8} \bigl( \chi^{2}_{1} + \chi^{2}_{2} + \chi^{2}_{3} + (v+\psi)^{2} \bigr)^{2}  \nonumber \\ 
& & \quad 
+ \frac{\epsilon m^{2}_{ex}}{2}\bigl(  \chi^{2}_{1} + \chi^{2}_{2} - \chi^{2}_{3} - (v+\psi)^{2} \bigr),  \\
\tilde{\partial}_{\nu} &=& \partial_{\nu}-i\mu\delta_{0\nu}.
\end{eqnarray}
After using the expression of the VEV $v$ ( Eq.(84) ) of $\Phi$, 
we get the quadratic part of the Lagrangian in the following form:
\begin{eqnarray}
{\cal L}^{(2)} &=& \frac{1}{2}\bigl( \partial_{\nu}\psi\partial^{\nu}\psi + \partial_{\nu}\chi_{a}\partial^{\nu}\chi_{a} \bigr)  \nonumber \\
& & 
+ \mu\bigl( \chi_{1}\partial_{0}\chi_{2} - \chi_{2}\partial_{0}\chi_{1} 
+ \chi_{3}\partial_{0}\psi - \psi\partial_{0}\chi_{3} \bigr)  \nonumber \\
& & + \epsilon m^{2}_{ex}(\chi^{2}_{1}+\chi^{2}_{2}) - (\mu^{2}+m^{2}_{0}-\epsilon m^{2}_{ex})\psi^{2}.
\end{eqnarray}  
$\epsilon > 0$ must be excluded to avoid tachyonic modes in $(\chi_{1},\chi_{2})$,
and this condition holds for the dispersion relations after diagonalization of ${\cal L}^{(2)}$ ( see, (92) ) 
Note that the total Lagrangian ${\cal L}$ is a class function ${\cal L}(\Phi)={\cal L}(g\Phi g^{-1})$, $g\in G$,
while both the effective potential $V_{eff}(\Phi=\Phi_{0})$ and ${\cal L}^{(2)}(\delta\Phi=\Phi-\Phi_{0})$ are not class functions,
even if the explicit symmetry breaking parameter vanishes.
This fact is crucially important for us to understand the physics of our anomalous NG theorem.
In the case of normal NG theorem, $V_{eff}(\Phi=\Phi_{0})$ is "effectively" a constant under the action of $g\in G$ since it does not contain any
coordinate of the NG manifold explicitly.
While, in our anomalous NG theorem, $V_{eff}(\Phi=\Phi_{0})$ will acquire an energy=mass by the action of $g\in G$,
and the energy which corresponds to the mass of an NG boson is provided from ${\cal L}^{(2)}(\delta\Phi)$.
Therefore, the effective potential of the theory is certainly periodically modulated under the action of $g\in G$ if $G$ is compact.
We will see this is the case in our discussion given below.
By using the path-integral formalism $\int{\cal D}\Phi{\cal D}\Phi^{\dagger}e^{i\int{\cal L}(\Phi,\Phi^{\dagger})}$,
we immediately recognize that this phenomenon of anomalous NG theorem is a pure quantum effect, 
since the Lagrangian of the beginning, ${\cal L}(\Phi,\Phi^{\dagger})$, 
is also a class function of any group operation $g\in G$ when any explicit symmetry breaking parameter vanishes. 
Namely, it is impossible to understand it by a tree-level, 
and we need at least the one-loop level to obtain the finite curvature along with a massive NG-bosonic coordinate.
This is a remarkable fact since the treatment of the symmetry breaking of the traditional bosonic Goldstone model 
( and also the Standard Model Higgs sector )
is understandable at the tree level, while a fermion composite model such as an NJL-type model has to evaluate at least 
a one-loop effective potential.    
To make the Lagrangian in a Hermitian matrix form explicitly,
we perform partial integrations in time-derivatives and rearrange the Lagrangian:
\begin{equation}
{\cal L}^{(2)} = \frac{1}{2}\tilde{\Phi}\left(
\begin{array}{cccc}
k^{2}+\epsilon m^{2}_{ex} & ik_{0}\mu & 0 & 0 \\
-ik_{0}\mu & k^{2}+\epsilon m^{2}_{ex} & 0 & 0 \\
0 & 0 & k^{2} & ik_{0}\mu \\
0 & 0 & -ik_{0}\mu & k^{2} -M^{2} 
\end{array}
\right)\tilde{\Phi},  
\end{equation}
where,
\begin{eqnarray}
\tilde{\Phi} &=& (\chi_{1},\chi_{2},\chi_{3},\psi)^{T},  \\
M^{2} &=& \mu^{2} + m^{2}_{0} - \epsilon m^{2}_{ex}.
\end{eqnarray}
Then we get
\begin{eqnarray}
E^{\chi_{1},\chi_{2}}_{\pm} &=& 
\sqrt{{\bf k}^{2}-\epsilon m^{2}_{ex}+\frac{\mu^{2}}{2}\pm\frac{\mu}{2}\sqrt{\mu^{2}+4{\bf k}^{2}-4\epsilon m^{2}_{ex}}},  \\
E^{\chi_{3},\psi}_{\pm} &=&
\sqrt{{\bf k}^{2}+\frac{M^{2}+\mu^{2}}{2}\pm\frac{1}{2}\sqrt{4\mu^{2}{\bf k}^{2}+(M^{2}+\mu^{2})^{2}}}.
\end{eqnarray}
Now it is clear for us from these dispersion relations at the limit ${\bf k}\to 0$ and $\epsilon\to 0$, 
$E^{\chi_{1},\chi_{2}}_{+}$ and $E^{\chi_{3},\psi}_{+}$ are massive while $E^{\chi_{1},\chi_{2}}_{-}$ and $E^{\chi_{3},\psi}_{-}$ are massless.

Therefore, we find that the masses of NG bosons in the $SU(2)$-Higgs-Kibble-type model 
are coming from the mode-mode coupling of a pair of broken generators.
This fact provides us a confirmation on our general discussion given in the previous section.
Hence, $\langle[Q^{1},Q^{2}]\rangle\ne 0$ and other commutators must have vanishing VEVs in this case.
This is achieved by the form of the vacuum $\Phi_{0}=(0,v)/\sqrt{2}$.
It must be distinguished that the commutators $[Q^{A},Q^{B}]$ are given by conserved charges and now 
$\langle Q^{1}\rangle=\langle Q^{2}\rangle=0$, $\langle Q^{3}\rangle\ne 0$,
while an order parameter can take the form 
$\Phi\sim a_{1}\sigma^{1}+a_{2}\sigma^{2}+a_{3}\sigma^{3}$, $a_{1}\ne 0$, $a_{2}\ne 0$, $a_{3}\ne 0$.
A Heisenberg algebra is obtained, and a pairwise decoupling takes place.
In this example of kaon condensation, the amplitude mode $\psi$ is defined toward the direction of VEV, 
and it couples with $\chi_{3}$. 
This amplitude mode might be expressed more generally such as $\psi \to \exp\psi((1-\sigma^{3})/2)\Phi_{0}$,
though this is not an element of the Lie group $SU(2)$, rather a projection operator, 
and then the manner of mode-mode coupling between $\psi$ and $\chi_{3}$ is not given in the same manner of commutator $[Q^{1},Q^{2}]$,
different from the mode-mode coupling of $\chi_{1}$ and $\chi_{2}$.
It should be noticed that 
$\chi_{1}=\Re\delta\phi_{1}$, $\chi_{2}=\Im\delta\phi_{1}$,
$\psi=\Re\delta\phi_{2}$, $\chi_{3}=\Im\delta\phi_{2}$,
where $(\delta\phi_{1},\delta\phi_{2})$ are fluctuations in the vicinity of the stationary point.
Thus, the mode-mode couplings relevant for our anomalous NG theorem take place between the real and imaginary parts
of $\phi_{1}$ and $\phi_{2}$ separately.
This fact in our kaon condensation model is quite interesting toward a classification of several possible types of mode-mode couplings
in NG bosons. In the above example, the mode-mode couplings of the NG bosons and the amplitude 
are described over a two independent discs of Gaussian plane.
It may be possible to generate a mode-mode coupling which gives a finite mass to an NG boson via a radiative correction:
This can be regarded as a kind of Coleman-Weinberg mechanism in an NG boson mass matrix.
This can be understandable if there is an interaction between two NG bosons ( such as an electromagnetic interaction ):
Especially, a Rayleigh-Schr\"{o}dinger or a quasi-degenerate perturbation theory can apply to the case where the spectrum
of an NG sector is split by small but finite masses.

The periodicity of the NG sector inside the Lagrangian can be understood as follows.
Since the massive NG bosons arises from the pair $(\chi^{1},\chi^{2})$, we prepare
\begin{equation}
g = e^{i(\chi_{1}\sigma^{1}+\chi_{2}\sigma^{2})} = \left(
\begin{array}{cc}
\cos|\chi| & i\frac{\chi_{-}}{|\chi|}\sin|\chi| \\
i\frac{\chi_{+}}{|\chi|}\sin|\chi| & \cos|\chi| 
\end{array}
\right),
\end{equation} 
where,
\begin{eqnarray}
|\chi| &=& \sqrt{\chi^{2}_{1}+\chi^{2}_{2}}, \quad  \chi_{\pm} = \chi_{1} \pm i\chi_{2}.
\end{eqnarray}
Then we evaluate
\begin{eqnarray}
\tilde{\partial}_{\nu}g\Phi_{0} &=& i\Bigl[(\tilde{\partial}_{\nu}\chi_{1})\sigma^{1}+(\tilde{\partial}_{\nu}\chi_{2})\sigma^{2}\Bigr]g\Phi_{0},
\end{eqnarray}
and take the following inner product, we get
\begin{eqnarray}
& & \tilde{\partial}^{\dagger}_{\nu}(\Phi_{0}g^{-1})\cdot\tilde{\partial}_{\nu}(g\Phi_{0})    \nonumber \\
& & \quad = v^{2}\Bigl[ 
\tilde{\partial}^{\dagger}_{\nu}\chi_{1}\tilde{\partial}_{\nu}\chi_{1} +
\tilde{\partial}^{\dagger}_{\nu}\chi_{2}\tilde{\partial}_{\nu}\chi_{2} + 
i\bigl( \tilde{\partial}^{\dagger}_{\nu}\chi_{1}\tilde{\partial}_{\nu}\chi_{2}
- \tilde{\partial}^{\dagger}_{\nu}\chi_{2}\tilde{\partial}_{\nu}\chi_{1} \bigr)\cos 2|\chi| \Bigr].
\end{eqnarray}
Therefore, the chemical potential $\mu$ acquires a periodic modulation proportional to the trigonometric function 
such that $\sim \mu^{2}\cos 2|\chi|$, and thus the mass of the dispersion $E^{\chi_{1},\chi_{2}}_{+}$ becomes periodic as a function of $|\chi|$.
Namely, the periodic modulation of the effective potential is kinematically generated in our anomalous NG theorem 
by the mode-mode-coupling caused by a finite chemical potential:
This is a quite remarkable result, since our result explains both the mechanism of generation of a finite mass of an NG bosons,
and a kinematically generated periodicity of the effective potential beyond the tree-level defined over the $SU(2)$ group manifold. 
It should be mentioned that this expression of the kinetic part of the Lagrangian is obtained by choosing a specific form of the VEV of $\Phi$,
i.e., $\langle\Phi\rangle = \Phi_{0}=(0,v)^{T}/\sqrt{2}$,
and thus this expression is a "function" of the form of VEV, the special form of local coordinates $|\chi|$, and the chemical potential $\mu$.
Hence, if we choose another type of VEV to $\Phi$, then in fact we will obtain another expression different from (97):
From this sense, both $V^{(tree)}_{eff}$ and ${\cal L}^{(2)}$ are not class functions of $SU(2)$
( caution: the kinetic term given above contains all of the orders of fluctuations $(\chi_{1},\chi_{2})$, not only ${\cal L}^{(2)}$ ).
The form $|\chi|$ implies that the theory is isotropic toward the directions $\chi_{1}$ and $\chi_{2}$ ( axial symmetric ),
similar to the case of a ferromagnet. 
One should notice that the Lagrangian of (88) or (89) is defined locally, at a specific point over the $SU(2)$ Lie group manifold.
To see the periodicity, we need a group element which is defined globally, as we have used above.
Now we obtain the effective potential of the model as $V_{eff}\sim V^{(tree)}_{eff}+ f(v)\mu^{2}\cos 2|\chi|$ 
by using the result of $E^{\chi_{1},\chi_{2}}_{+}$
( $f(v)$: a scalar function of the VEV $v$ ), 
which shows a periodicity toward the direction of "amplitude" $|\chi|=\sqrt{\chi^{2}_{1}+\chi^{2}_{2}}$,
while it is flat to the direction of the phase ( precession mode ) of $\chi_{1}+i\chi_{2}$ 
( the amplitude and the phase defines an infinite number of $S^{1}$ circles of a Gaussian plane ), as we have stated above.
This fact is parallel with the case of ferromagnet, 
where the massless NG mode ( spin wave ) is the precession described by a linear combination of 
the two modes $(\sigma_{1},\sigma_{2})$.
Absolutely interesting fact we have found here is that this global structure of the effective potential
( periodic toward the radial direction $|\chi|$ while exactly flat along with the phase variable )
is coming from the uncertainty relation arises from the Heisenberg algebra obtained from $SU(2)$:
There is a strong uncertainly toward the phase direction, 
while the motion toward the radial direction is well localized and the "position" is determined by the set of periodic stationary points:
The set of stationary points gives a Galois symmetry. 
In the vicinity of a stationary point ( valley ), a representation point has a small fluctuation toward the radial direction
while it strongly fluctuate along with the phase coordinate. 
Namely, 
\begin{flushleft}
{\bf Theorem}:
{\it The uncertainty relation of the Heisenberg algebra obtained from the Lie algebra of $SU(2)$
realizes in the global structure of the effective potential of a theory. 
One degree of freedom is almost fixed/determined while another degree of freedom of the Heisenberg pair of shows a strong uncertainty.}
\end{flushleft}  
We argue that a quasi-Heisenberg algebra generically obtained in various symmetry breaking schema of our anomalous NG theorem
will determine the structure of effective potential according to satisfy the uncertainty relations.
Therefore, a mass generation in an NG sector in our anomalous NG theorem reflects the uncertainty principle!
Furthermore, an explicit symmetry breaking mass parameter such as the prescription of explicitly+dynamical symmetry breakings
also acts to fix a phase degree of freedom, and thus, a mass spectrum any meson ( mesonic state ) in a non-Abelian Lie group symmetry 
is generically a result of uncertainty relation.

The kaon condensation model we consider here has a lot of similar aspects with physics of a ferromagnet.
Needless to say, the time-reversal symmetry is broken in a ferromagnet, 
and a precession of magnetization reflects this time-reversal symmetry breaking. 
The system of kaon condensation we discuss here might have an effect of time-reversal symmetry breaking,
caused by the NG bosons, at its low-energy excited state.

We make a brief comment on the Higgs ( Anderson-Higgs-Brout-Englert-Guralnik-Hagen-Kibble ) phenomenon in the Lagrangian (81)~[2,21,28,34].
If we put a $U(1)$ gauge field to the Lagrangian, it also gives a Higgs phenomenon with the field redefinition 
$U_{0}=\mu+A_{0}-\partial_{0}\theta$, $U_{i}=A_{i}-\partial_{i}\theta$ ( $\theta$: a $U(1)$ phase ) and thus they give a massive Proca theory.
While if we put a set of $SU(2)$ gauge fields as (81), 
the chemical potential $\mu$ plays no role and the usual Higgs phenomenon is observed.

\section{The Model Lagrangian Approach}

In all of the above classification of types of NG bosons given in the end of the introduction of this paper, 
the dispersion relations and mass spectrum of NG bosons should be obtained from an analysis of effective action $V_{eff}$.
In general, an effective action and/or a potential are expanded by bosonic fields, 
thus we restrict ourselves to the case of bosonic ( bosonized ) theory.
Since the NG boson Lagrangian will be obtained from $V_{eff}$, 
we construct a Lagrangian to make our problem more tractable. 
Let ${\cal L}(\Phi)$ be a Lagrangian, and let us take a small displacement of bosonic field as $\Phi=\Phi_{0}+\delta\Phi$.
$\Phi$ is assumed to belong to a representation of $G$.
$\delta\Phi$ contains the NG bosons and the amplitude mode. 
In the case of $SU(N)$, one frequently use a fundamental representation, ${\bf N}\ni\Phi, \Phi_{0}, \delta\Phi$.
Let us consider the case of symmetry given by a Lie group $G$ with ${\rm dim}{\rm Lie}(G)=N$.
Then ${\rm dim}(\delta\Phi)=N+1$, where the additional one degree of freedom is the amplitude mode of $\Phi$.
Then the Lagrangian is expanded into the following form:
\begin{eqnarray}
{\cal L}(\Phi_{0}+\delta\Phi) &=& {\cal L}(\Phi_{0}) + \frac{\partial{\cal L}(\Phi_{0})}{\partial(\delta\Phi)}\delta\Phi 
+ \frac{1}{2!}\frac{\partial^{2}{\cal L}(\Phi_{0})}{\partial(\delta\Phi)^{2}}(\delta\Phi)^{2} + \cdots.
\end{eqnarray}  
The first-order derivative vanishes in the effective action, and the second-order derivative gives the mass matrix and dispersion relations of
NG bosons. Namely,
\begin{eqnarray}
\frac{\partial^{2}{\cal L}(\Phi_{0})}{\partial(\delta\Phi)^{2}} &=& \Delta^{-1}_{F}(p_{\nu}).
\end{eqnarray}
This equation corresponds to (70).
The algebraic roots of ${\rm det}\Delta^{-1}_{F}(p_{\nu})=0$ gives the dispersion relations of NG bosons.

One can also consider the case where the order parameter is a vector/tensor $\Phi_{\mu\nu\cdots\rho}$,
explicitly breaks the Lorentz symmetry. 
In that case, we can consider the following formal expansion:
\begin{eqnarray}
{\cal L}(\Phi_{\mu\nu\cdots\rho}) &=& \sum^{\infty}_{n=0}\frac{1}{n!}\frac{\partial^{n}{\cal L}}{\partial(\delta\Phi_{\mu\nu\cdots\rho})^{n}}
(\delta\Phi_{\mu\nu\cdots\rho})^{n}.
\end{eqnarray}
Here, we do not consider contractions of the Lorentz indices $(\mu,\nu,\cdots,\rho)$.
A vectorial order parameter is frequently found in superconductivity or $^{3}$He superfluidity~[50,65,66,78].
The second-order derivative term as the quadratic part of quantum fluctuations,
\begin{eqnarray}
{\cal L}^{(2)} &=& \frac{1}{2}\frac{\partial^{2}{\cal L}}{\partial(\delta\Phi_{\mu\nu\cdots\rho})^{2}}(\delta\Phi_{\mu\nu\cdots\rho})^{2}
\end{eqnarray}
gives the mass matrix and dispersion relations of NG bosons.
While, due to the Coleman-Mandula theorem, a Poincar\'{e}-invariant theory can have only a conserved charge of scalar type.
We currently consider a Lorentz-violating system of (non)relativistic theory, and thus we assume a theory can have vector/tensor-charges
$Q^{A}_{\mu\nu\cdots\rho}$ of internal symmetries.
Hence, formally,
\begin{eqnarray}
\delta^{A}_{\mu\nu\cdots\rho}\Phi^{A'}_{\mu'\nu'\cdots\rho'} &=& [Q^{A}_{\mu\nu\cdots\rho},\Phi^{A'}_{\mu'\nu'\cdots\rho'}] 
\end{eqnarray}
will be considered.
Hence, we find that the Lie brackets $[Q^{A}_{\mu\nu\cdots\rho},\Phi^{A'}_{\mu'\nu'\cdots\rho'}]$ define how ${\cal L}^{(2)}$
is given in terms of the NG bosons, 
similar to the case of $V_{eff}$ we have discussed in the previous section. 
Namely, we will consider the Lie brackets of internal symmetries with Lorentz indices.
Thus, a quasi-Heisenberg algebra should be obtained from $[Q^{A}_{\mu\nu\cdots\rho},Q^{B}_{\mu'\nu'\cdots\rho'}]$ 
in a diagonal symmetry breaking scheme.
If this algebra causes a mode-mode coupling in quantum fluctuations of NG bosons, 
then our anomalous NG theorem takes place.
Hence, we argue it is enough for us to consider a scalar field to study the mechanism of our anomalous NG theorem.

Now, we can systematically construct a generic Lagrangian which may show the phenomenon of anomalous NG theorem.
We know from our observation on the model of kaon condensation, the relevant part of the Lagrangian
to give the anomalous behavior of NG bosons is its kinetic term.
For example, in the case of two-component real bosonic field:
\begin{equation}
{\cal L} = \frac{1}{2}(\phi_{1},\phi_{2})\left(
\begin{array}{cc}
\partial_{\nu}\partial^{\nu} & a_{\nu}\partial^{\nu}  \\
a^{*}_{\nu}\partial^{\nu}  & \partial_{\nu}\partial^{\nu}
\end{array}
\right)\left(
\begin{array}{c}
\phi_{1}  \\
\phi_{2}
\end{array}
\right).
\end{equation} 
The off-diagonal part describes a mode-mode coupling:
From our examination on the Lagrangian of NG sector of the kaon condensation model, 
it is now obvious fact that the explicit symmetry breaking {\it mass} parameter enters into the diagonal part of the Lagrangian matrix,
while the mode-mode coupling matrix elements take of course several off-diagonal elements. 
We do not consider kinetic terms of derivatives higher than the second-order,
since they may cause a tachyonic mode or a negative norm state. 
Therefore, this mechanism of anomalous NG theorem must break the Lorentz symmetry.

With respect to the general theory of nonlinear sigma models, the following Lagrangian is examined:
\begin{eqnarray}
{\cal L} &=& \frac{1}{2}\delta\Phi{\cal M}\delta\Phi   \nonumber \\
&=& \frac{1}{2}g_{ab}\partial_{\nu}(\Psi)^{a}\partial^{\nu}(\Psi)^{b} -\frac{1}{2}\Psi M^{2}\Psi +  
\Psi\left( 
\begin{array}{ccc}
0 &  & \sum\tilde{a}_{\nu}\partial^{\nu}  \\
  &  \ddots &  \\
\sum\tilde{a}^{\dagger}_{\nu}\partial^{\nu} & & 0
\end{array}
\right)\Psi,   \\
\Psi &=& (\chi_{1},\cdots,\chi_{N},\phi)^{T}, \quad ( a= 1,2,\cdots,N ).
\end{eqnarray} 
Here, the real matrix $g_{ab}$, which defines a Riemannian metric $\hat{g}=g_{ab}dx^{a}\otimes dx^{b}$ ( $dx^{a},dx^{b}\in{\bf R}$), 
must satisfy a condition given by the Lie group $G$.
Usually, it is taken to be a unit matrix, i.e., 
which defines a Euclidean space, a typical example of a simply connected Riemannian symmetric space, 
with its dimension equal to the number of broken generators.
It is known fact that, in a small displacement $\Psi \to \Psi + \delta\Psi$, $\delta\Psi$ 
must be a Killing vector defined over the target space of a nonlinear sigma model.
Here, we have put explicit symmetry breaking mass parameters $M^{2}$ ( a Hermitian matrix ). 
$\tilde{a}_{\nu}$ are matrices, and they can contain both explicit symmetry breaking parameters ( like a chemical potential )
or spontaneously generated parameters from its underlying theory.
Therefore, the number of roots of
\begin{eqnarray}
{\rm det}{\cal M}(p_{\nu} \ne 0) &=& 0
\end{eqnarray}
gives the number of NG bosons plus 1.
From the mathematical structure of ${\cal M}$, we find that the matrix can be expressed by a Borel subalgebra 
( given by the direct sum of Cartan subalgebra ${\bf h}$ and the positive-weight part of ${\bf g}$ ) of ${\rm Lie}GL(N+1)$ or ${\rm Lie}O(N+1)$~[43].
Probably, we can employ a flag variety to study a representation of such an expression of Borel subalgebra, Borel subgroup $B$, the Borel-Weil theorem, 
and a Bruhat decomposition of $G=\cup_{\omega\in W}B\omega B \ni g$ as a disjoint union
( $G$; a connected reductive algebraic group, $W$; a Weyl group ):
The mode-mode coupling part of the matrix ${\cal M}$ is decomposed into a strictly upper triangular matrix and a strictly lower triangular matrix,
and they can be expressed by a Borel subalgbera. 
( Note: 
Any strictly triangular matrix is nilpotent, and a set of strictly upper/lower triangular matrices forms a nilpotent Lie algebra. 
Borel subgroup = invertible triangular matrices = all diagonal entries must be nonzero. 
Borel subalgebra = not necessarily invertible. )
Of particular interest is a homogeneous space $G/P=BWB/P$ ( $P$; a parabolic subgroup ) which gives a parabolic geometry.
Since ${\cal L}\in {\rm Lie} O(N+1)$ and the local coordinate system ( NG bosons ) are all real, 
one can consider an adjoint group action $g^{-1}{\cal L}g$ ( $g\in O(N+1)$ ). 
The equation ${\rm det}{\cal M}(p_{\nu} \ne 0)=0$ defines a hypersurface in such a Borel subgroup representation space.
The Borel-Weil theorem states that the global section $\Gamma(G/B,L_{\lambda})$ 
( $L_{\lambda}$ is a $G$-equivariant holomorphic line bundle over $G/B$, $\lambda$ denotes a dominant integral highest weight )
gives an irreducible representation of $G$.
Hence, the mechanism of generating massive NG bosons reflects a breaking of $O(N+1)$ ( more precisely, $O(N+1,{\bf R})$ ) symmetry
as a result of a breaking scheme and a Lorentz-violating parameter,
and the breaking structure of $O(N+1)$ reflects the dispersion relations and numbers of massive/massless NG bosons.
We will see later that this $O(N+1)$ structure gives a symmetry ( in fact, degeneracy ) of the mass spectrum of NG bosons:
In fact, as we will see at (116) of the decomposition into several Heisenberg pairs,
the number and structure of Heisenberg pairs embedded in ${\cal L}$ is determined by the $O(N+1)$ structure.
When we observe our Lagrangian from the context of chiral perturbation theory, 
the expansion of a chiral perturbation may be given in a form of a series of symplectic matrices
( a series of the form proportional to $\sum_{n}c_{n}(1+c_{A}T^{A})^{n}$, $T^{A}\in {\rm Lie}(G)$ ).

Our result and discussion given here depends on our observation of the kaon condensation model with the general theory
of effective action/potential, and thus we cannot say definitely whether there is another mechanism
( different from the mode-mode coupling mechanism ) which gives a similar phenomenon/result
of anomalous NG theorem at a Lagrangian level.
Our result given here is derived from a relativistic model, though it can be applied to the cases of (anti)ferromagnets,
since their low energy excitations are described by the class of $O(3)$ ( and mathematically generally, $O(N)$ ) nonlinear sigma models.
In that case, according to our examination of the Heisenberg algebra coming from $SU(2)$,
mode-mode coupling terms similar to the case of relativistic model with a finite chemical potential may be introduced in the sigma model of a nonrelativistic case.
Therefore, if such type of nonlinear sigma models can be derived from Ginzburg-Landau-Goldstone-Higgs-Kibble-type theories,
then similar phenomenon may be observed.
In fact, the Lagrangian of ferromagnet given in the paper of Watanabe and Murayama~[82] 
takes almost a special example of our generic Lagrangian (104).
Namely, 
\begin{flushleft}
{\bf Theorem}: {\it The mechanism of anomalous NG theorem in a nonrelativistic case is the same with that of a Lorentz-violating relativistic case. The counting law of the number of NG bosons of a nonrelativistic case is the same with that of a Lorentz-violating relativistic case}.
\end{flushleft}
It is interesting for us to study a phase of magnon condensation by our theoretical framework presented here,
since a condition of Bose-Einstein condensation is determined by a mass parameter and a chemical potential
( a magnon energy in a real substance is typically $\mu$eV ).

Our generic Lagrangian given above can be discussed more systematically and mathematically/geometrically~[42].
With respect to the procedure to obtain the quadratic part in terms of NG bosons from the Lagrangian of kaon condensation,
a kinetic part is expressed as follows:
\begin{eqnarray}
\Phi &=& g\Phi_{0}, \qquad g = e^{iQ^{A}\chi^{A}} \in G, \\
\tilde{\partial}^{\dagger}_{\nu}\Phi^{\dagger}\tilde{\partial}^{\nu}\Phi &=&
-\Phi^{\dagger}_{0}(g^{-1}\tilde{\partial}_{\nu}g)(g^{-1}\tilde{\partial}^{\nu}g)\Phi_{0},    \nonumber \\
&=& -\Phi^{\dagger}_{0}\Bigl(g^{-1}(\partial_{\nu}\partial^{\nu}-2i\mu\partial_{0}-\mu^{2})g\Bigr)\Phi_{0}.
\end{eqnarray} 
Here, we have chosen the local coordinate system $\{\chi^{A}\}$ as the first kind.
The term of $2i\mu\partial_{0}$ is coming from the Leibniz rule of derivation.
In general, the kinetic term is given by 
$g_{\alpha\bar{\beta}}\tilde{\partial}^{\dagger}_{\nu}(\Phi^{\dagger})^{\beta}\tilde{\partial}^{\nu}(\Phi)^{\alpha}$,
and $\tilde{g}=g_{\alpha\bar{\beta}}dz^{\alpha}\otimes d\bar{z}^{\beta}$ ( $dz^{\alpha}\in{\bf C}$, $d\bar{z}^{\beta}\in \bar{\bf C}$ ) 
gives a Hermitian metric. 
This Lagrangian does not have the fluctuation of amplitude mode of $\Phi$, 
which can couple with an NG boson as we have observed in the kaon condensation model. 
In fact, the amplitude mode cannot be expressed by a Maurer-Cartan form.
It should be noticed that, 
\begin{eqnarray}
g^{-1}\tilde{\partial}_{0}g = g^{-1}\partial_{0}g -i\mu.
\end{eqnarray}
Note that this equation shows a "deformation" of ( or, a deviation from ) the Maurer-Cartan 1-form 
\begin{eqnarray}
\omega(T^{A},\theta^{A})=g^{-1}\partial_{\nu}g= i\sum T^{A}\otimes\theta^{A}, \quad \theta^{A} = \partial_{\nu}\chi^{A}, 
\end{eqnarray}
as a geometric object
( $\{T^{A}\}$ give the Maurer-Cartan frame, $\{\theta^{A}\}$ are the Maurer-Cartan coframe ),
or a deviation from the Killing form given in terms of $g^{-1}\partial_{\nu}g$, 
namely ${\rm tr}(g^{-1}\partial_{\nu}g,g^{-1}\partial^{\nu}g)$ in the Lagrangian.
It must be noticed that $g^{-1}\partial_{0}g\in {\rm Lie}(G)\simeq T_{e}G$ ( $e$: the origin ),
while the part $-i\mu$ is independent from a geometric structure of ${\rm Lie}(G)$.
A Maurer-Cartan form is an adjoint orbit, defines essentially a homogeneous space.
Moreover, a Maurer-Cartan form defines a local section:
A vacuum state of the quantum field theory is given by the specific local section of $V_{eff}$ in the sense of fiber bundle,
and a structural group ( namely, a Lie group ) gives a transformation between two local sections. 
The effective potential $V_{eff}$ is an example of so-called representation function:
$V_{eff}$ is a continuous function defined over a topological space $X$,
and has a continuous group action of a group element of $G$.
Thus, the group orbit is defined by the pair $(V_{eff},g\in G)$, which is a subset of the space of continuous functions over $X$.
In other words, $V_{eff}$ belongs to a set of continuous sections of a $G$-bundle $E$.
Namely, $V_{eff}\in \Gamma(E)_{G}\subset \Gamma(E)$, where $\Gamma(E)$ is the space of all continuous sections,
$\Gamma(E)_{G}$ is its submodule.
$\Gamma(E)_{G}$ is dense in $\Gamma(E)$ when $G$ is compact.
This fact is important to certify a variational calculus of $V_{eff}$ to obtain a stationary point.
Note that the Killing form is now subjected to the Euler-Lagrange variation principle, 
and of course defines a phase factor of path integral.
In the usual case, a connection $\xi$ is introduced by the form $g^{-1}(d+\xi)g$, and the part $g^{-1}dg$ is the Maurer-Cartan form,
namely the chemical potential is included in our theory under the manner of a connection.
Note that in our case, $\chi^{A}$ have the additional dependence on spacetime coordinates coming from
the local "wave" nature of quantum field theory which is not contained in the traditional Lie theory.
From our results, we know the mode-mode couplings of NG bosons are given as VEVs of the space of Lie algebra of the 1-form,
such as $g^{-1}\mu(\partial_{0}\chi^{A})Q^{A}g$, namely a displacement at the origin caused by a Lie group in the derivative of NG boson,  
and the explicit symmetry breaking parameter $\mu$ acts on them as a linear scaling factor. 
Moreover, any deformation theory of mathematics of manifolds defined over a field of characteristic zero
is described by Maurer-Cartan elements of a differential graded Lie algebra 
( for example, the theory of deformation quantization of Kontsevich~[45] ).
In our case, since our anomalous NG theorem quite often gives a "quasi" Heisenberg algebra over a Poisson manifold, 
from locally to globally, thus, we need a special study on a deformation theory and a deformation quantization
which can achieve a quasi-Heisenberg algebra/group.   
In addition, our $\omega$ satisfies the Maurer-Cartan structural equation, sometimes called as the deformation equation,
\begin{eqnarray}
& & d\omega(T^{A},\theta^{A}) + \frac{1}{2}[\omega(T^{A},\theta^{A}),\omega(T^{A},\theta^{A})] = 0, \quad {\rm or},   \\
& & d\theta^{A} + i \sum_{B,C}f^{BCA}\theta^{B}\wedge\theta^{C} =0.
\end{eqnarray}
The Maurer-Cartan equation is the vanishing condition of the curvature 2-form, namely
the vanishing condition of the curvature of Cartan connection $\Omega=d\omega+\frac{1}{2}[\omega,\omega]=0$.
( Understood as Maurer-Cartan form $\subset$ Cartan connection. )
The Maurer-Cartan equation always holds for any $\omega$.
This definition of curvature is independent from the notion of "curvature" 
we use in the second-order derivative of the effective potential $V_{eff}$:
In fact, $V_{eff}$ is a function of local coordinates, while a transition function between two local coordinate systems over $V_{eff}$ 
is determined by the geometric structure of $V_{eff}$ itself, thus the transition function cannot be defined group theoretically {\it a priori} 
( mostly, $GL(n,{\bf R})$ ).
The symmetry in the vicinity of a point defined over $V_{eff}$ can be found from the curvature matrix as the second order derivative of $V_{eff}$
since the mass matrix of NG sector reflects the symmetry at a point.
If the mass spectrum of NG sector has a degeneracy, then the symmetry at the point becomes higher.  
Now the Cartan connection is an affine connection of frame bundle ( i.e., a tangent bundle ) of the base manifold,
and also be interpreted as a special example of connection of principal bundle.
Due to the Cartan-Ambrose-Hicks theorem, a manifold is locally Riemannian symmetric
if and only if its curvature is constant, and for any simply connected, complete locally symmetric space is Riemannian symmetric. 
The expressions of a Lagrangian given in terms of $\omega= g^{-1}dg$ 
are familiar in theory of nonlinear realization~[49]: The reason is needless to say.
For a case of a Clifford-Klein form $\Gamma\backslash G/H$ of a symmetric space, 
the following parametrization will be introduced:
\begin{eqnarray}
\Phi &=& \gamma gh\Phi_{0}, \quad \gamma \in \Gamma, \, g\in G, \, h\in H.
\end{eqnarray}
Hence, a lattice element $\gamma$ is contained in the Maurer-Cartan form $(\gamma gh)^{-1}d(\gamma gh)$.
Here, as we have mentioned above, a lattice $\gamma$ gives a symmetry of a set of stationary points of $V_{eff}$ over $G/H$,
and thus $\gamma$ implies an equivalence of physics of $(gh)^{-1}d(gh)$ and $(\gamma gh)^{-1}d(\gamma gh)$
in an NG boson Lagrangian ( a kind of replication of a theory takes place ).
Due to involutions and $\gamma$, the global structure of the NG manifold in a symmetric space is well-characterized.
The part of mode-mode couplings in our Lagrangian is coming from the cross terms of 
a product of similarity transformations in the above expression (108).
In fact, these expressions have the advantage to consider some geometric nature of the anomalous NG theorem,
since the manner of how a Lorentz-symmetry-violating parameter couples with an internal symmetry of Lie group
will be explained by the words of geometry.
Note that $\partial_{\nu}$ are generators of subalgebra of Poincar\'{e} algebra.
It should be mentioned that, when $\Phi$ is an order-parameter of Dirac mass type which spontaneously breaks a chiral $\gamma_{5}$ symmetry 
( coming from an underlying fermion system ), a transform $\Phi\to e^{2i\gamma_{5}\theta_{5}}\Phi$ can be considered in the Lagrangian,
though it does not give a mode-mode coupling between NG bosonic fields inside the Lagrangian. 
Hence, at least in our definition of bosonic Lagrangian, the structure of mode-mode couplings of NG bosons we have revealed in this paper
is closed inside the internal symmetry of a Lie group. 
( Or, can say an internal symmetry can couple with a Poincar\'{e} generator $\partial_{\nu}$ but cannot couple with the $\gamma_{5}$-rotation. )
On the other hand, there are couplings between the chiral $\gamma_{5}$-symmetry
and NG bosons in an explicit+dynamical symmetry breaking in an NJL-type $SU(N)$ model with $N>2$ ( see~[70] ).
In the latter case, the $\gamma_{5}$-chiral symmetry and the flavor symmetry are coupled with each other through the chiral projection operator
$P_{\pm}=\frac{1\pm\gamma_{5}}{2}$.
In Ref.~[70], the author discussed that a chiral mass can be handled as a Riemann surface.
Since a $\gamma_{5}$-transformation is vertical with the internal symmetry of $SU(2)$, $\Phi$ gives a direct sum of Riemann surfaces.
( In the case of $SU(N)$ with the case when $\Phi$ takes a fundamental representation ${\bf N}$ or $\overline{\bf N}$,
a direct sum of $N$ Riemann surfaces will be given.  )
In such a case, since some NG bosons become massive and the effective potential $V_{eff}$ is periodic with respect to the directions
of massive modes in our anomalous NG theorem, some Galois-group symmetries of both $\gamma_{5}$ and massive modes arise simultaneously in a theory.
In other words, the theory gives a Galois representation of those degrees of freedom ( $\gamma_{5}$, massive modes ) simultaneously.

Toward the understanding of counting law of massive NG bosons, we will examine the kinetic term of bosonic Lagrangian given above,
especially by using the words of Cartan geometry. 
The relevant part of mode-mode couplings of bosonic fields is
obtained from the Lie-algebra expansion of the Maurer-Cartan 1-form $g^{-1}\partial_{0}g$, namely, 
\begin{eqnarray}
\Phi^{\dagger}_{0}\Bigl\{e^{-i\sum_{A}Q^{A}\chi^{A}}2i\mu \sum_{B}Q^{B}(\partial_{0}\chi^{B})e^{i\sum_{A}Q^{A}\chi^{A}}\Bigr\}\Phi_{0} 
\end{eqnarray}
Here we have used the fact that $\Phi_{0}$ is a constant ( VEV ), has no dependence on spacetime coordinates.
This term can arise only by the presence of the Lorentz-symmetry-violating parameter $\mu$ in the Lagrangian.
After expanding the exponential mappings and picking up the quadratic terms of bosonic fields, we get
\begin{eqnarray}
(114) &=& -2\mu\sum_{A}\sum_{B}\chi^{A}(\partial_{0}\chi^{B})\Phi^{\dagger}_{0}[Q^{B},Q^{A}]\Phi_{0}   \nonumber \\
&=& 2\mu{\sum\sum}_{A>B}\Bigl\{ \chi^{A}(\partial_{0}\chi^{B})-\chi^{B}(\partial_{0}\chi^{A})\Bigr\}\Phi^{\dagger}_{0}[Q^{A},Q^{B}]\Phi_{0}.
\end{eqnarray}
This is a sum of off-diagonal part ( upper triangular part without the diagonal elements ) of matrix elements with indices $A,B$.
The number of non-vanishing terms of this sum, namely the part 
$\sum_{A>B}\Phi^{\dagger}_{0}[Q^{A},Q^{B}]\Phi_{0} = i\sum_{A>B}f^{ABC}\Phi^{\dagger}_{0}Q^{C}\Phi_{0}$,  
counts the number of pairs of two modes they are coupled with each other. 
Maximally the number of terms is $N(N-1)/2$ when $N={\rm dim}{\rm Lie}(G)$,
and thus the number coincides with the number of generators of $SO(N)$:
Hence the linear space of mode-mode coupling matrix can be expanded by the basis set of Lie$(SO(N))$.
Moreover, each term of this sum has a correspondence with the Lie algebra valued linear equations 
given in our discussion by using the second-order derivatives of the effective potential $V_{eff}$, namely Eqs. (68), (72) and (75).
The symplectic structure discussed in Ref.~[81] is obvious in our expression (115) due to the anti-symmetric nature
of structural constants $f^{ABC}$.
Thus, our Lagrangian defines a symplectic manifold with an appropriately defined symplectic structure.
A linear transformation in the matrix space of our Lagrangian gives an isomorphism of the symplectic structure, possiblly continuously.
Moreover, if a symplectic transformation of our Lagrangian is given over a symplectic manifold, then it is expressed as a symplectic Lie group. 
For example, in the case of $SU(2)\to U(1)$ of a ferromagnet with broken generators $(Q^{1},Q^{2})$ and a symmetric generator $Q^{3}$,
only $if^{123}\Phi^{\dagger}_{0}Q^{3}\Phi_{0}$ remains to give a finite VEV. 
If we choose the representation that gives $Q^{3}$ in the form of diagonal matrix such as the Pauli matrix $\sigma^{3}$, 
then this term remains when the VEV of $\Phi$ takes the form $\Phi_{0}=(v_{1},v_{2})$ ( $v_{1}\ne v_{2}$ ).
In this case, the number of pair of bosonic fields they are coupled is 1. 
When the VEVs of $[Q^{A},Q^{B}]$ are pairwise decoupled, then the set of VEVs gives a set of Heisenberg algebras,
and then the mode-mode couplings in the space of NG bosons $(\chi_{1},\cdots,\chi_{N})$ are also decoupled into subspaces pairwisely,
and the problem of the matrix is reduced into the direct sum of $2\times 2$ matrices ( i.e., block-diagonalized ) such that
\begin{eqnarray}
\left( 
\begin{array}{cc}
k^{2} & 2ic_{1}\mu k_{0} \\
-2ic_{1}\mu k_{0} & k^{2}
\end{array}
\right)\oplus\cdots\oplus\left(
\begin{array}{cc}
k^{2} & 2ic_{l}\mu k_{0} \\
-2ic_{l}\mu k_{0} & k^{2}
\end{array}
\right).
\end{eqnarray}
In such a case, our discussion on dispersion relations of NG bosons is reduced very much.
The decomposition to $2\times 2$ matrices can also be interpreted as the result that a Lie algebra is constructed by
the fundamental unit $sl_{2}$-triple, apparent from a Cartan decomposition.  
This is the origin of the counting law of Watanabe, Brauner, and Hidaka:
It is obvious from our analysis, rank$\langle [X^{A},X^{B}] \rangle$ counts the number of pairs of mode-mode coupling,
and thus it can estimate the dimension of a matrix of non-vanishing mode-mode coupling elements, 
though the pairwise decoupling must take place to conclude definitely that the number of massive NG bosons is {\it the half} of the rank. 
The cases of 
\begin{eqnarray}
& & \langle h_{i}\rangle\ne 0, \quad \langle e_{j}\rangle\ne 0, \quad \langle f_{k}\rangle= 0,
\end{eqnarray}
and
\begin{eqnarray}
& & \langle h_{i}\rangle\ne 0, \quad \langle e_{j}\rangle\ne 0, \quad \langle f_{k}\rangle\ne 0,
\end{eqnarray}
give more complicated situation for the counting law.

The adjoint orbit $O(X)={\rm Ad}(G)X$ ( $X\in{\rm Lie}(G)$ ) defines a submanifold of ${\rm Lie}(G)$.
A typical example is the Maurer-Cartan form $g^{-1}dg$, 
and the curvature 2-form will be defines as a function of the adjoint orbit.
Needless to say, an NG manifold consists of adjoint orbits:
\begin{eqnarray}
O(\Phi) &=& g^{-1}\Phi g \simeq e^{i\chi^{A} X^{A}}\Phi, \quad \Phi \in {\rm Lie}(G).
\end{eqnarray}
After choosing the specific form of $\Phi\in {\rm Lie}(G)$
( for example, when $\Phi$ takes its value in the Cartan subalgebra of ${\rm Lie}(G)$ ), 
$O(\Phi)=g^{-1}\Phi g$ gives a homogeneous space $G/G(\Phi)$, where $G(\Phi)$ is the stabilizer 
$\{g\in G| {\rm Ad}(g)\Phi=\Phi\}$. 
If $G$ is compact, an adjoint orbit is called as an elliptic orbit.

Especially the case of $SU(2)\to U(1)$ gives a more explicit geometric interpretation.
Let us consider a situation where a one-dimensional curve is defined over a two-dimensional oriented surface,
and the surface is embedded into a Euclidean space ${\bf R}^{3}$.
Let ${\bf T}$ be the unit tangent vector of the curve, let ${\bf N}$ be the unit normal vector of the surface, 
and ${\bf S=N\times T}$ is the tangent normal.
The ${\bf T}$, ${\bf N}$ and ${\bf S}$ gives the Darboux frame, an orthogonal frame.
Let $C_{normal}$ be the normal curvature, $C_{geodesic}$ be the geodesic curvature, and let $T_{geodesic}$ be the geodesic torsion. 
It is a known fact that those quantities are given by the following linear transformation ( the Frenet-Serret equation ):
\begin{equation}
\frac{d}{ds}\left( 
\begin{array}{c}
{\bf T} \\
{\bf S} \\
{\bf N}
\end{array}
\right) = \left( 
\begin{array}{ccc}
0 & C_{geodesic} & C_{normal}  \\
-C_{geodesic} & 0 & T_{geodesic}  \\
-C_{normal} & -T_{geodesic} & 0
\end{array}
\right)\left( 
\begin{array}{c}
{\bf T} \\
{\bf S} \\
{\bf N}
\end{array}
\right).
\end{equation}
Apparently, the matrix of this linear transformation is a group element of $SO(3)$.
Hence the off-diagonal part of our Lagrangian of the anomalous NG theorem in the $SU(2)$ case has this geometric implication,
a curve on a surface in ${\bf R}^{3}$. 
An important difference is that ${\bf S}$ is defined by ${\bf T}$ and ${\bf N}$,
while $(\chi_{1},\chi_{2},\chi_{3})$ are linearly independent with each other:
Namely, the projective case $\chi^{2}_{1}+\chi^{2}_{2}+\chi^{2}_{3}=1$ corresponds to the Frenet-Serret equation.
A matrix element of the off-diagonal part corresponds to a curvature or a torsion,
and the local coordinates of $SU(2)$ ( i.e., the NG bosons ) gives an orthonormal frame.
In other words, $[Q^{A},Q^{B}]$ define the local geometry of the NG manifold.
Those mathematical structure may be hidden in the back ground of physics of a ferromagnet.  
Hence, in a case of pairwise decoupling, only one of $C_{geodesic}$, $C_{normal}$, and $T_{geodesic}$ remains finite,
which indicates that the three dimensional space $(\chi_{1}.\chi_{2},\chi_{3})$ is decomposed into a one- and a two-dimensional spaces,
and the mixing of them along with the curve only takes place in the two-dimensional subspace,
and the one-dimensional subspace is inert. 
Our interpretation on the Lagrangian of NG sector in $SU(2)$ is quite natural and not surprising one,
since the cross terms of the kinetic part $(g^{-1}(d+\xi)g)(g^{-1}(d+\xi)g)$ contains the off-diagonal elements of
the Lagrangian matrix, and the chemical potential $\mu$ acts like a connection $\xi$ inside the Lagrangian.
More explicitly, the NG bosons $(\chi_{1},\chi_{2},\chi_{3})$ form an orthogonal local coordinate system of the $SU(2)$ group manifold,
and the "curvatures" and "torsions" reflect the geometric effect on the local coordinates $(\chi_{1},\chi_{2},\chi_{3})$ 
displaced by the Lagrangian. 
Namely, they measure how the curve generated by a collective motion of NG bosons are distorted.
This mathematical/physical interpretation of our Lagrangian is more clarified if the Lagrangian formalism
is converted into a Hamiltonian of equations of motion.
Hence, the dynamical equation of NG bosons itself keeps this geometric nature.
This interpretation of the geometric implication of our NG-bosonic Lagrangian can apparently be applied to more general case, 
for example, $SU(N)$.
Namely the off-diagonal matrix elements proportional to chemical potential $\mu$, which give the mode-mode couplings between
NG bosons and cause massive spectra of them, act as curvatures/torsions to the local coordinate system ( i.e., the NG bosons )
of $SU(N)$ Lie group manifold.
In other words, the chemical potential $\mu$ gives a measure of how much the group manifold 
( more precisely, the NG manifold as the submanifold of $SU(N)$ ) have the finite curvatures/torsions.
Therefore, the kinetic part ${\cal L}_{K}$ can be rewritten symbolically as
\begin{eqnarray}
{\cal L}_{K} &=& \frac{1}{2}{\rm tr}\Phi\bigl( (g^{-1}dg)^{2} +\mu \Omega_{a}T^{a}  \bigr)\Phi,
\end{eqnarray}
where, $\{T_{a}\}$ denote the Lie algebra of orthogonal group, $\Omega_{a}$ indicate curvature 2-forms.
The NG bosons give a Darboux frame in general, an example of moving frame, which is closely related with the Maurer-Cartan form.
It should be noticed that an NG bosons as a Darboux frame is always holds, no matter the case of normal or anomalous NG theorem.
While a curvature matrix vanishes in a normal case, and it takes finite matrix elements in an anomalous case.
The Cartan's method of moving frames is applied to study the local structure of a homogeneous space $G/H$:
Hence now we find something about the local nature of an NG manifold of the anomalous NG theorem.
If we employ a normalization condition of the vector $(\chi_{1},\cdots,\chi_{N})$ 
( namely a unit vector $\chi^{2}_{1}+\cdots+\chi^{2}_{N}=1$ ),
with taking a special orthogonal group for the algebra $T^{a}$ of the above Lagrangian, 
then it contains an isometry group, may be expressed as a Killing vector field.
It is a known fact that Killing vector fields form a Lie algebra. 
Moreover, a set of Killing vector fields is related with a curvature tensor.

The part of mode-mode coupling terms for NG bosons caused by the chemical potential $\mu$ in our Lagrangian, (104) or (108),
is not pairwisely decoupled in general, and thus it does not show a Heisenberg algebra apparently.
At least, via VEVs $\Phi^{\dagger}[Q^{A},Q^{B}]\Phi$, a subspace of the $N$-dimensional space must be decomposed such as
$(1,2)\oplus(3,4)\oplus\cdots\oplus(M-1,M)$, ( $M < N$ must be satisfied ) to show a Heisenberg algebra.
Thus, we cannot conclude the VEVs of Eq. (115) are always Heisenberg-type, depend on cases and breaking schema. 
While, due to the Hermitian nature of Eq. (115) and any non-vanishing off-diagonal matrix element in the momentum space takes pure-imaginary,
and the number of independent matrix elements are $N(N-1)/2$, the matrix, namely the mode-mode coupling part of our Lagrangian
can be expressed by the generators $L_{j}$ of $SO(N)$ ( angular momenta of ${\rm Lie}(SO(N))$, $j=1,\cdots, N(N-1)/2$ ):
\begin{eqnarray}
(\tilde{\partial}_{\nu}\Phi)^{\dagger}(\tilde{\partial}_{\nu}\Phi) &=& -\frac{1}{2}\Psi\Bigg\{ {\rm diag}(\partial^{2}_{\nu}, \cdots, \partial^{2}_{\nu}, \partial^{2}_{\nu}-M^{2}) + i\sum^{N(N-1)/2}_{j=1}c_{j}L_{j} \Bigg\}\Psi + \cdots,
\end{eqnarray}
( $c_{j}$; coefficients ).
These angular momenta $L_{j}$ in an $N$-dimensional real Euclidean space ( locally ${\bf R}^{N}$ ) 
describe rotations of coordinates, namely the NG bosons, on the group manifold $G$.
Since a quantum mechanical mixing of NG bosons takes place in our theory of anomalous NG theorem,
it is quite interesting that those NG bosons may give a multiplet structure in their energy spectrum!
Hence it might be possible to introduce a weight space ${\cal V} = \oplus_{\lambda}{\cal V}_{\lambda}$,
starting from the highest weight. 
In a case of ferromagnet, the highest weight state of Lie$(SU(2))$ is the eigenstate of the Heisenberg Hamiltonian,
and $S_{\pm}=S_{1}\pm iS_{1}$ provides the ladder operators.
A similar situation can take place in a Lie$(SU(N))$ model under our anomalous NG theorem.

Since the chemical potential $\mu$ takes a similar form with a zeroth component of gauge field $A_{0}(x)$,
we speculate a similar situation takes place when the NG sector couples with gauge fields.
Let us give a general theory for understanding this situation.
Let us write a generic differential operator $\cal{D}_{\nu}$, 
which gives a covariant derivative in the sense of gauge theory as its special case, and make a similarity transformation:
\begin{eqnarray}
{\cal D}_{\nu}(\{Q^{A}\}) &=& \partial_{\nu}(\{Q^{A}\}) + \delta{\cal D}_{\nu}(\{Q^{A}\}),  \\
\partial_{\nu}(\{Q^{A}\}) &=& g^{-1}\partial_{\nu}g,  \\
\delta{\cal D}_{\nu}(\{Q^{A}\}) &=& g^{-1}{\cal B}g,  \\
g &=&  e^{iQ^{A}\chi_{A}} \in G, \\
{\cal B} &=& {\cal B}^{0}\hat{1} + {\cal B}^{\alpha}\tau^{\alpha} = B + B_{\nu} + B_{\nu\mu} + B_{\nu\mu\rho} + \cdots.
\end{eqnarray}
Here, the part of $\delta{\cal D}_{\nu}(\{Q^{A}\})$ denote the "displacement" from the Maurer-Cartan 1-form 
caused by some explicit symmetry breaking parameters or gauge fields.
The Lie algebra $\tau^{\alpha}$ in which the gauge fields $B_{\nu}$ take their values are in principle different from
( no relation with ) the broken generator $Q^{A}$.
The matrix ${\cal B}$ is considered as a set of Lorentz-symmetry-violating parameters with various tensors.
We do not consider any gravitational effect but it can be incorporated.
Then the kinetic term is assumed to take the following expression defined over a bosonic field $\Phi$,
and one can expand it:
\begin{eqnarray}
& & ({\cal D}_{\nu}(\{Q^{A}\})\Phi)^{\dagger}({\cal D}^{\nu}(\{Q^{A}\})\Phi)   \nonumber \\
& & \quad = \Phi^{\dagger}\Bigg\{ -\partial_{\nu}(\{Q^{A}\})\partial^{\nu}(\{Q^{A}\})
-\partial_{\nu}(\{Q^{A}\})\delta{\cal D}_{\nu}(\{Q^{A}\})  \nonumber \\
& & \qquad +\delta{\cal D}_{\nu}(\{Q^{A}\})^{\dagger}\partial_{\nu}(\{Q^{A}\}) 
+\delta{\cal D}_{\nu}(\{Q^{A}\})^{\dagger}\delta{\cal D}_{\nu}(\{Q^{A}\}) \Bigg\}\Phi.
\end{eqnarray}
The mode-mode couplings between bosonic fields including the NG bosons are coming from
\begin{eqnarray}
\Phi^{\dagger}\Bigl[ -\bigl( \delta{\cal D}_{\nu}(\{Q^{A}\}) - \delta{\cal D}_{\nu}(\{Q^{A}\})^{\dagger} \bigr) \partial_{\nu}(\{Q^{A}\})
-\bigl\{ \partial_{\nu}(\{Q^{A}\})\delta{\cal D}_{\nu}(\{Q^{A}\}) \bigr\} \Bigr]\Phi. & & 
\end{eqnarray}
If we restrict ${\cal B}$ as vector components ( connection ), 
then the expression of the kinetic term is reduced into the following form by using the Maurer-Cartan 1-form:
\begin{eqnarray}
{\cal L}_{K} &=& \frac{1}{2}\Phi\Bigl( (g^{-1}(d+A)g)(g^{-1}(d+A)g) \Bigr)\Phi   \nonumber \\
&=& \frac{1}{2}\Phi\Bigl( g^{-1}(\partial\cdot\partial + \partial\cdot A + A\cdot\partial + A^{2})g\Bigr)\Phi.
\end{eqnarray}
Since the group element of broken symmetry $g$ has no relation with the gauge field $A$,
and thus $A$ can be taken as a scalar matrix proportional to a unit matrix ( namely, an electromagnetic field ) for our present purpose. 
Then $A$ gives a similar effect with $\mu$ inside the Lagrangian:
This is a remarkable result since this can be experimentally confirmed,
and it might be possible to confirm our anomalous NG theorem by mesons of QCD under an electromagnetic field
( external, static, constant, modulated ).
One should notice that ${\cal L}^{(2)}$ is manifestly gauge invariant. 
For example, if $A=(A_{0}={\rm const}., 0,0,0)$, then it gives the same situation with a finite $\mu$.
Therefore, the chemical potential $\mu$ plays a similar role with an external magnetic field in a spin system:
Again, we have met with a phenomenological similarity between our anomalous NG theorem and an explicit symmetry breaking.

Since NG bosons can have finite masses in our anomalous NG theorem,
they have finite ranges for their propagations similar to the Yukawa pion.
( The correlation length is finite to the direction of a massless mode, while diverges toward a massive mode. )
Therefore, their corresponding orderings in a matter may be of short ranges:
A long-range ordering observed by an experiment realizes via a remaining massless NG boson.
Since the range of propagation of a massive NG boson is shorter than a massless one,
the interaction between amplitude-mode particles mediated by a massive NG bosons also has a finite range. 
It indicates that there is an anisotropy in an ordered state in a matter.
In the case of a ferromagnet, an off-diagonal element causes a mixing between $\chi^{1}$ and $\chi^{2}$ modes 
with the same weight, and then it results a massive and a massless modes.
Thus, there is no anisotropy toward the $x$ and $y$ directions,
the axial symmetry around the $z$-axis is kept.

Let us consider an $SU(4)$ spin-orbital model which has been studied in condensed matter physics~[41], 
with assuming a "ferromagnetic" ordering takes place.
In addition, the diagonal breaking, namely, all Lie algebra generators except the Cartan subalgebra are broken is assumed.
Thus, a conserved charge $Q^{A}$ which belongs to the Cartan subalgebra taking a nonvanishing VEV.
Now rank(Lie$(SU(4))$) is 3, while dim(Lie$(SU(4))$) is 15, then the number of Heisenberg-algebra-like pair is (15-3)/2=6.
Furthermore, if the low-energy Lagrangian of this system takes the similar form 
with the kaon condensation model with the pairwise decomposition like (89) or (116),
then the ferromagnetic state has 6 massless and 6 massive NG bosons: 
Each mass eigenstate belongs to a sextet, would be called as NG-boson multiplets. 
If each Lagrangian of 6 pairs are the same form, then an $SO(6)\otimes SO(6)$ symmetry arises in the spectrum of the theory.
It should be noticed that the number of generators of this group is 15+15=30, larger than that of $SU(4)$.
The reason is that the Ginzburg-Landau-Goldstone-Higgs-Kibble-type $SU(4)$ Lagrangian is given in terms of a complex scalar field,
while the NG bosons are given by the real and imaginary part of the complex field.
At the massless limit, the mass spectrum has $SO(12)$ symmetry ( now the dimension of Lie$(SO(12))$ is 66 ).
Since a quasi-Heisenberg relation will be obtained in this case, 
we predict a modification of the Heisenberg uncertainty relation may be observed in the NG sector of this $SU(4)$ spin-orbital model. 
Hence,
\begin{flushleft}
{\bf Theorem}:
{\it A diagonal breaking of $SU(N)$ under the situation of anomalous NG theorem discussed here
gives the $SO((N^{2}-N)/2)\otimes SO((N^{2}-N)/2)$ symmetry ( massive and massless ) in the mass spectrum of NG bosons.
At the massless limit, the spectrum of NG bosons has the $SO(N^{2}-N)$ symmetry.
Thus, phenomenologically, the anomalous NG theorem is understood as a symmetry breaking 
$SO(N^{2}-N)\to SO((N^{2}-N)/2)\otimes SO((N^{2}-N)/2)$.
The $SO(M)$ symmetry of the NG manifold means the isotropy of an effective potential $V_{eff}$ at a point of the NG manifold
where a Lie algebra is defined and examined ( local ).
The $SO(M)$ itself can be used globally as a structural group of vector bundle constructed by NG boson fields.}
\end{flushleft}

It is noteworthy to mention that the author observed an $SO(2)$ symmetry arises clearly
in the case of pseudo-NG bosons of the flavor symmetry breaking of $SU(2)\to U(1)$
in an explicit+dynamical symmetry breaking of an NJL-type four fermion model~[70].
The $SO(M)$ symmetry arised from an $SU(N)$ model is already discussed by in Ref.~[70].
The appearance of orthogonal Lie group symmetry in an NG-boson sector is not yet examined enough in spite of its importance.

The importance of orthogonal group symmetry in a mass spectrum of NG bosons in our anomalous NG theorem is 
that they are coming from the eigenvalues of geometric curvature matrix.
In our case of the anomalous NG theorem, the effective potential depends on the local coordinates ( i.e., the NG bosons ) of a Lie group,
and thus the potential has a nonvanishing curvature in general: 
The curvature reflects the dependence of $V_{eff}$ on those coordinates. 
The global behaviors of those curvatures will have the correspondence with the Lie group manifold and the NG manifold as its subspace,
though, several local properties of them must be distinguished.
If the mass spectrum of the NG sector has a symmetry ( degeneracy ) such as an orthogonal group discussed above,
then the curvature matrix of the effective potential has a symmetry. 
In general, an NG sector has a degeneracy in the mass spectrum, whether the situation is normal or anomalous.

By using a four-dimensional Heisenberg-type spin model, some similarity between anomalous and explicit symmetry breakings would be understood.
A Heisenberg-like spin model ${\cal H}_{spin}$ is obtained via a Killing form:
\begin{eqnarray}
{\cal H}_{spin} \sim {\cal J}{\rm tr}(S^{a},S^{a}) \sim {\rm tr}({\rm Ad}(G),{\rm Ad}(G)) \simeq {\rm tr}(g^{-1}dg,g^{-1}dg),
\end{eqnarray}
( where, $S^{a}\in {\rm Lie}(G)$, and ${\cal J}$ implies an isotropic coupling constant with respect to the indices of the Lie algebra ).
Then, with including an external field or a small perturbation parallel with a ferromagnetic mean field 
( which is taken to the third direction in the following form )
which acts as a term of an explicit symmetry breaking parameter, the Hamiltonian is 
\begin{eqnarray}
\widetilde{\cal H} &\sim& {\cal J}{\rm tr}(g^{-1}dg,g^{-1}dg) + {\cal H}_{ex+mf},  \\
{\cal H}_{ex+mf} &\propto& S^{3} \sim (g^{-1}dg)_{3} + {\rm h.c.} 
\end{eqnarray} 
( h.c. means the Hermitian conjugate ).
${\cal H}_{ex+mf}$ contains both the contributions of an external field and a mean ( "molecular" ) field.
After taking a derivative expansion of spacetime coordinates, we yield a nonlinear sigma model of ferromagnet defined over $G$:
This observation is similar with the case of explicit symmetry breaking. 
From the form of ${\cal H}_{ex+mf}$, it is apparent for us that ${\cal H}_{ex+mf}$ gives 
a term which may have a similar role of the chemical potential discussed by our model Lagrangian in this section.
It is interesting for us to consider the case when $G$ is an exceptional Lie group,
from the perspective of geometric property of our anomalous NG theorem.

Usually, the Langevin equation formalism is utilized as the canonical approach to study an irreversible process and a dynamical critical phenomenon.
The Langevin equation ( a stochastic differential equation~[71,74] ) is given as a first-order differential equation, 
which has its theoretical back ground in theory of nonrelativistic Brownian motions.
The diffusion equation which will be obtained at the long-wave-length/hydrodynamic limit of a Langevin equation,
is also a nonrelativistic equation, of course:
We never have met with a relativistic diffusion equation which might belongs to the world of hydrodynamics.
( Recently, an attempt toward a theory of relativistic Brownian motions and a relativistic Langevin equation has been published~[19].
Mathematically, we need a framework of relativistic stochastic differential equation and relativistic Ito diffusion. )
Thus, currently there is an essential difficulty to adopt our anomalous NG theorem to
those nonrelativistic theoretical frameworks.
In other words, a difference between relativistic and nonrelativistic cases of our anomalous NG theorem might be found in some problems of dynamics.

In summary, a violation of Lorentz symmetry by a certain mechanism ( explicitly or spontaneously ) 
in a Lagrangian causes a modification of its low-energy effective theory which describes NG bosons,
then the subset of NG bosons acquires masses. 
Probably, a certain type of deformation of a sigma model Lagrangian can generically give a massive mode.
Our Lagrangian can also be generalized to supersymmetric nonlinear sigma models of several types:
In such a model, a massive NG fermion might appear simultaneously with a massive NG boson.
A supersymmetric theory frequently used in particle phenomenology has a usual Lie group/algebra,
thus it may be the case that we will consider a usual Lie group/algebra ( not Lie supergroup/superalgebra )
to investigate an anomalous behavior of NG theorem.
For examples of supersymmetric field theory with finite chemical potentials, see~[68,69]:
Our anomalous NG theorem can be extended to SUSY cases via the results of these references.
To find and establish the counting law for SUSY cases is an important subject for particle phenomenology.

\subsection{Poincar\'{e}, Conformal, Super-Poincar\'{e}, Superconformal Groups and Some Lie Groups in the Anomalous NG Theorem}

Until now, we examine the anomalous NG theorem by the following logic:
(1) The Lorentz symmetry is broken in a theory, (2) then a special coupling between elements of a Lie algebra of internal symmetry is caused
via the Lagrangian, (3) then a violation of the normal NG theorem takes place.
We try to extend this logic to (1') Poincar\'{e}, conformal, super-Poincar\'{e}, or superconformal symmetries are broken in a theory,
(2') then some couplings between Lie algebras of an internal symmetry are caused inside the Lagrangian of a theory,
(3') then a violation of the normal NG theorem takes place.

At least a formal discussion is quite easy. 
Let us consider the largest case, a superconformal group, its superconformal algebra, 
and a ( semisimple ) Lie group of internal symmetry of a theory.
Then let us assume a Lagrangian in which some generators of the superconformal algebra are broken spontaneously/explicitly by a VEV or 
an explicit symmetry breaking parameter. Then the Lagrangian is assumed to have a mode-mode coupling term of the Lie algebra of internal symmetry
via the explicit symmetry breaking parameter.
From this logic, it is clear for us that a Lie bracket which will be examined for studying our anomalous NG theorem 
belongs to the Lie algebra of internal symmetry.  
Let ${\cal L}$ a be Lagrangian, 
and let ${\cal Q}^{A}$ ( $A=1,\cdots,S$ ) be the N\"{o}ther charges of internal symmetries associated with Lie groups, 
and $j^{A}$ the corresponding conserved N\"{o}ther currents.
Then we add $j^{A}$ to ${\cal L}$ as a Legendre transform:
\begin{eqnarray} 
{\cal L}(\Phi,\Phi^{\dagger}) -\sum^{S}_{A=1}\mu^{A}j^{A}(\Phi,\Phi^{\dagger}).
\end{eqnarray}
The multipliers $\mu^{A}$ are explicit symmetry breaking parameters ( of a superconformal group ),
conjugates of conserved currents $j^{A}$.  
Then we obtain the NG boson Lagrangian of the quadratic part:
\begin{eqnarray}
{\cal L}^{(2)} \sim \frac{1}{2}{\rm tr}\Phi_{0}\Bigg\{ (g^{-1}dg)^{2} - \sum^{S}_{A=1}\mu^{A}\langle\frac{\delta^{2}j^{A}}{\delta\Phi^{2}} \rangle \Bigg\}\Phi_{0} + \cdots.
\end{eqnarray}
The second term inside the curly bracket may cause model-mode couplings between NG bosons.
$\langle\cdots\rangle$ indicates a VEV.
Since an analysis on the algebraic structure of mode-mode couplings of NG bosons is examined by VEVs of Lie algebra of an internal symmetry,
certainly a quasi-Heisenberg algebra arises also in a (super)conformal/Poincar\'{e}-violating case.

\section{The Riemann Hypothesis and the Nambu-Goldstone Theorem: Toward the Solution}

Here, we discuss an interesting aspect of mathematical implication of the Nambu-Goldstone theorem
to the Riemann hypothesis, the Bost-Connes model~[7], and class field theory~[54].
In fact, when we consider an explicit+dynamical symmetry breaking~[70], 
the mathematical structure of the NG theorem acquires a viewpoint closely related 
with the mechanism of the phenomena of the Riemann hypothesis. 
This fact implies us a natural solution=proof on the Riemann hypothesis, which has been unsolved 154 years, 
might be found along with the direction of the mathematical structure of the NG theorem.
( In this paper, we do not discuss a possible way toward the solution of the Riemann hypothesis,
which remains for our future efforts. ) 
Since an explicit+dynamical symmetry breaking and our anomalous NG theorem share some similarities,
we consider here this problem.

In the paper of Connes and Marcolli~[14], 
they discussed that the physical implication of some algebra of the Bost-Connes model is understood by a phase factor 
( which takes a similar form to a coherent state representation ) which takes its form as 
the $N$-th root of unity.
Besides the ordinary Bost-Connes model, the "generated" cyclotomic field associated with a spontaneous symmetry breaking 
should take place in quantum field theory, i.e., a system of an infinite number of dynamical degrees of freedom.
A quantum field theory is usually defined over ${\bf R}^{n}$ or ${\bf C}^{n}$ with some quantum numbers associated with symmetries of the theory,
while a cyclotomic field is a Galois extension of ${\bf Q}$.
This fact implies that a model which generates a Galois group "effectively" changes a number field
via a certain mechanism or a functor,
associated with a change of topology and cardinality of a number field as a base space of the system.
This phenomenon is quite often observed also in the NG theorem, both its generalization~[70] and our anomalous NG theorem:
This is the starting point of our discussion toward the mechanism of the phenomena of the Riemann hypothesis.
In our NG theorem, we can consider a coset $({\bf Z}/N{\bf Z})\backslash G$, and the bosonic field is given by
\begin{eqnarray}
\Phi &=& \zeta_{N}g\Phi_{0}, \quad \zeta_{N} = e^{2\pi i/N}, \, g\in G.
\end{eqnarray}
Then the Lagrangian will be constructed by the formalism of nonlinear realization~[49].
In this case, the Galois group symmetry, a cyclotomic extension, is introduced implicitly 
in the kinetic part ( the Killing form ) of the Lagrangian,
while the potential/mass term may contain the Galois symmetry explicitly.
The $\zeta_{N}$ as a phase factor of a wavefunction will vanish inside the Maurer-Cartan form and the Killing form 
( since of course $\zeta_{N}$ does not have a spacetime dependence ).
Therefore, when $\zeta_{N}$ is introduced to a theory explicitly, the theory acquires something beyond the framework
of Cartan geometry constructed by the 1-form $g^{-1}dg$.
It is noteworthy to mention that the quantity $\zeta_{N}=e^{2\pi i/N}$ takes its value in a unit circle of ${\bf C}$,
while ${\bf Z}/N{\bf Z}$ is arised as a symmetry defined by the quantity.
This quite simple fact indicates us that the current issue certainly occupy its place in the 
mechanism of spontaneous symmetry breaking.
Moreover, the symmetry ${\bf Z}/N{\bf Z}$ is the symmetry of several vacua given by a theory ( as argued in the ordinary Bost-Connes model ),
while it cannot be an NG bosonic mode if we keep ourselves inside the ordinary NG theorem:
The statement of ordinary/normal NG theorem translated by the words of effective potential is that a spontaneous symmetry breaking
gives a flat direction toward a local coordinate of broken generator of a Lie group.
In other words, all points on the NG manifold are equivalent.
While, such a discrete symmetry can be obtained via the generalized NG theorem~[70],
or our anomalous NG theorem with a breaking of equivalence between points of an NG manifold. 
Of course, we can introduce $({\bf Z}/N{\bf Z})\backslash G$ as the structural group of a fiber bundle 
( for example, defined by a Higgs field ) of a theory.

Let $M$ be a homogeneous space. Then a study on the $k$-rational points in $M$ is the problem of Galois cohomology
( $k$: a field ).
In general, a cohomology of group studies a set of fixed points under group actions.
Let $A$ be an Abelian module, and assume $Gal(K/k)$ acts on $A$.
Then the Galois cohomology group is 
\begin{eqnarray}
H^{n}(Gal(K/k),A), \quad n\ge 0.
\end{eqnarray} 
It is defined by the complex $(C^{n},d)$, where $C^{n}$ consists with all maps $Gal(K/k)^{n}\to A$,
and $d$ is the coboundary operator.
If $A$ is a non-Abelian case, only the zero-dimensional $H^{0}$ and one-dimensional $H^{1}$ cohomology can be defined.
In that case, $H^{0}(Gal(K/k),A)=A^{Gal(K/k)}$ is the set of fixed points under the action of $Gal(K/k)$ in A:
Thus, the invariant set $A^{Gal(K/k)}$ gives a representation of the Galois symmetry.
For example, in an explicit+dynamical symmetry breaking of a $U(1)$ group,
an embedding $A^{Gal({\bf Q}(\zeta_{N})/{\bf Q})} \to A^{U(1)}$ takes place
by the set of stationary points obtained from $V_{eff}$~[70].
This type of symmetry will be discussed later, in our discussion on the relation between the NG theorem 
and the Bost-Connes model, and the Riemann hypothesis.
In fact, the generalized NG theorem of explicit+dynamical symmetry breaking given by the author in Ref~[70]
has some examples where the ground state of a quantum field theory spontaneously acquires a Galois symmetry. 
In such a case, a set of discrete vacua arises from the symmetry breaking, 
and they are in fact the invariant subset $A^{Gal(K/k)}$ embedded in $A$.
Here, $A$ is the NG manifold expanded by the broken local coordinates coming from a subspace of the group manifold.
Hence, the set of discrete vacua of the generalized NG theorem gives a Galois representation.
Similar situation is realized in our anomalous NG theorem, since it gives a massive NG mode,
and the effective potential is lifted along with the local coordinate ( i.e., the NG boson )
then the effective potential must be periodic in the direction of local coordinate if the Lie group is compact:
{\it This is a dynamical mechanism for generating a Galois representation in quantum field theory.}
If we take a Maurer-Cartan form from the group element $\zeta_{N}g$, 
then $\zeta_{N}$ will be canceled inside the Maurer-Cartan form. 
Thus, a curvature 2-form derived from the Maurer-Cartan form, 
a characteristic class~[59] evaluated from the 2-form, and also the kinetic part of the Lagrangian, i.e., a Killing form,
cannot contain any information of the Galois symmetry. 
This fact implies that it is difficult to express a Galois symmetry by the modern differential-geometric setting.
While the mass ( potential energy ) term of a Lagrangian can explicitly give a Galois symmetry such that
\begin{eqnarray}
V(\zeta_{N},g) &\propto& \zeta_{N}g\Phi + \Phi^{\dagger}g^{\dagger}\zeta^{*}_{N}.
\end{eqnarray}

Let us show another perspective on the relation between the Bost-Connes model and our anomalous NG theorem.
The Hamiltonian $H_{BS}$ and the partition function $Z_{BS}$ of the Bost-Connes-type model are defined as follows:
\begin{eqnarray}
H_{BS} &=& \ln N,  \\
Z_{BS} &=& {\rm Tr}e^{-\beta H_{BS}},
\end{eqnarray}
( $\beta$; inverse temperature ). 
$N$ is the number operator of one-flavor bosonic field, and thus $H_{BS}$ is defined by a Heisenberg algebra.
Especially, $N$ can be regarded as ( a part of ) the second-order Casimir element 
( Laplacian, the center of the universal enveloping algebra ${\cal U}({\rm Lie}(G))$ ) of the Heisenberg algebra.
It is known fact from the result of Beilinson and Bernstein that there is a categorical correspondence
between the category of coherent $D$-modules and the category of finitely generated ${\cal U}({\rm Lie}(G))$-modules with a certain
condition given by the center of ${\cal U}({\rm Lie}(G))$~[5].
Thus, the representation problem of ${\rm Lie}(G)$ in our case discussed here can be translated to the problem of $D$-modules.
Let us consider, for example, our anomalous NG theorem of $SU(2)\to U(1)$ of a ferromagnet. 
Let $a$ be an annihilation operator of the Bost-Connes mode, and let $a^{\dagger}$ be its Hermitian conjugate.
Then, needless to say, we have the Heisenberg algebra $(a,a^{\dagger},c)$, $[a,a^{\dagger}]=c$, $[a,c]=[a^{\dagger},c]=0$.
By comparing this algebra with the Lie$(SU(2))$ algebra, we find/set the correspondence
$a\leftrightarrow S_{1}$, $a^{\dagger}\leftrightarrow S_{2}$, $c\leftrightarrow S_{3}$ from the context of our anomalous NG theorem,
generating a Heisenberg algebra from the Lie$(SU(2))$ algebra.
Therefore we find
\begin{eqnarray}
N \sim \frac{1}{2}(a^{\dagger}a + aa^{\dagger}) \simeq \frac{1}{2}(S_{1}S_{2}+S_{2}S_{1}).
\end{eqnarray}
Namely, $N$ is expressed in somewhat similar form of an $XY$-spin model ( $H_{XY}=\sum S_{x}(i)S_{x}(i\pm 1)+S_{y}(i)S_{y}(i\pm 1)$ ).
The algebraic/operator structure of $N$ given as a quadratic form of bosonic operators might be interpreted as a non-interacting bosonic system,
though we can say $(a,a^{\dagger})$ are given from a Hartree-Fock-Bogoliubov mean field theory.
The crucial point is that the Hamiltonian is diagonalizable 
against a Fock space, and the notion of occupation number is well-defined. 
From the context of our anomalous NG theorem, this form gives a mode-mode coupling of broken generators $(S_{1},S_{2})$
in the breaking scheme $SU(2)\to U(1)$ of a ferromagnet. 
We can generalize our statement.
Let ${\bf g}={\rm Lie}(G)$, and decompose it as ${\bf g}={\bf h}+{\bf m}={\bf h}\oplus_{\alpha\in R}{\bf g}_{\alpha}={\bf h}\oplus{\bf e}\oplus{\bf f}$.
Then, at least in a case of diagonal breaking scheme, we have the correspondence of Heisenberg and Lie algebras as follows:
\begin{eqnarray}
N &\sim& {\cal C}({\rm Lie}(G)),   \\
{\cal C}({\rm Lie}(G)) &\sim& \sum(e_{i}f_{i}+f_{i}e_{i}) \nonumber \\
&=& \sum (g_{\alpha}\otimes g_{-\alpha}+ g_{-\alpha}\otimes g_{\alpha})
\in {\rm tr}({\bf m}\otimes{\bf m})  \nonumber \\
&\sim& {\rm tr}\Bigl[(g^{-1}dg)_{\bf m}\otimes (g^{-1}dg)_{\bf m}\Bigr].
\end{eqnarray}
A Weyl group implicitly acts on the Casimir element, which will also be reflected to enforce a specification/restriction of 
the algebraic form of our interpretation of the Bost-Connes model.
One should notice that the part $\sum(e_{i}f_{i}+f_{i}e_{i})$ is just the Casimir element of the universal enveloping algebra of Lie$(G)$.
Thus, we can write
\begin{eqnarray}
\zeta(\beta) &=& {\rm Tr}e^{-\beta H} = {\rm Tr}e^{-\beta\ln {\cal C}({\rm Lie}(G))} = {\rm Tr}({\cal C}({\rm Lie}(G)))^{-\beta}. 
\end{eqnarray}
Here, $\zeta$ is the Riemann zeta function. 
Especially in the case where a symmetric space is spontaneously generated, we can write 
\begin{eqnarray}
H^{G/H}_{BS} &=& \ln \bigl[ {\rm tr}({\bf m}\otimes{\bf m})\bigr] \simeq \ln \bigl[ {\rm tr}(T_{e}(G/H)\otimes T_{e}(G/H) )\bigr],     \\
\zeta^{G/H}(\beta) &=& Z = {\rm Tr}\bigl[ {\rm tr}(T_{e}(G/H)\otimes T_{e}(G/H))\bigr]^{-\beta}.  
\end{eqnarray}
This might be understood as a generalization of Riemann zeta function.
Namely, it is given by a trace of direct product of adjoint orbits or tangent spaces. 
Therefore, our interpretation/generalization of the Bost-Conne-like model resembles with 
the notion of dynamical zeta function~[76], and also a chiral perturbation theory.
From our discussion given here, we can say a Riemann zeta function 
is a function of a sum of the number of quantum states caused by mode-mode couplings in our anomalous NG theorem.
How a Galois symmetry of cyclotomic extension will be found?
In the case $SU(2)\to U(1)$ of a ferromagnet of our anomalous NG theorem,
the ground state of $V_{eff}$ of the system is defined over a two-dimensional local coordinate system, 
where one is "massless" and $V_{eff}$ is flat along with this direction,
while another direction is "massive" and has a finite curvature, 
and $V_{eff}$ shows a periodicity along with the massive direction since $SU(2)$ is compact.
Then the set of discrete vacua gives a Galois symmetry, $Gal({\bf Q}(\zeta_{N})/{\bf Q})$.
Therefore, a Galois symmetry arises from the symmetry of several vacua of the theory, 
while our Bost-Connes-type model is evaluated as a kind of "invariant" or a "character" of the theory in our case:
This point is different from the ordinary Bost-Connes model, in which the Riemann zeta function arises as
the partition function of the model itself, and the cyclotomic Galois symmetry is the symmetry of the vacuum states of the model.
Since the effective potential $V_{eff}$ of the case $SU(2)\to U(1)$ of a ferromagnet gives a set of discrete vacua,
it defines a lattice of the generated Heisenberg algebra, ${\bf Z}\otimes X^{a}$
( $X^{a}$: the basis of Lie algebra ).
Our discussion is summarized as the following diagram:
\begin{flushleft}
Normal/generalized/anomalous NG theorem in quantum field theory $\to$ NG boson Lagrangian/Hamiltonian $\to$
Heisenberg algebra, residual symmetry between several vacua ( periodicity ) $\to$ 
Bost-Connes-type model, Galois symmetry $\to$ the Riemann zeta function.
\end{flushleft}
Our formalism of the Bost-Connes-type Hamiltonian by a Lie algebra can be extended to a case of Kac-Moody algebra.
( Someone might recall the Shintani-Witten zeta function from our result given above, but it is quite different. )
It can be stated that the Boltzmann factor $e^{-\beta H}$ is a kind of exponential mapping of the Heisenberg algebra:
Namely, the trace ( sum ) of the exponential mappings of the universal enveloping algebra of the Heisenberg algebra
with an appropriate Hilbert space gives the Riemann zeta function.
Thus, the Boltzmann factor $e^{-\beta H}$ is a kind of globalization ( analytic continuation ) of a Lie algebra ( a tangent space at the origin )
from a geometric point of view.
This simple observation is remarkable, since such a globalization can be achieved only by a non-compact Lie group,
to acquire the continuation of the whole part of Gaussian plane ${\bf C}$ from the perspective of the Riemann hypothesis.
( The set of zeroes of $\zeta$ is non-compact. )
Our result is summarized by the following diagram:
\begin{flushleft}
Lie algebra, or a central extension of symplectic algebra $\to$ quasi-Heisenberg algebra 
$\to$ Casimir element of the universal enveloping algebra
$\to$ continuation to the whole part of the Gaussian plane via the trace of exponentiation of the logarithmic function of the Casimir element
$\to$ the Riemann zeta function, the Riemann hypothesis.
\end{flushleft}
We will give the following theorem:
\begin{flushleft}
{\bf Theorem}: {\it A dynamical/spontaneous generation of a Heisenberg algebra of 
our anomalous NG theorem of quantum field theory gives a Riemann zeta function
via the prescription of the Bost-Connes model. 
A bosonic Fock space is associated with the Riemann zeta function automatically.}
\end{flushleft}

We would like to give some comments here.
The famous Deligne-Lusztig theory~[17] is defined for a finite reductive group under applying a Frobenius endomorphism,
and thus it is not exactly the same with an $p$-adic analog of local coordinates of a Lie algebra/group
sometimes obtained in our generalized/anomalous NG theorem.
For example, for $SL(n,{\bf K})$ ( ${\bf K}=\overline{\bf F}_{p}$ ), a Frobenius endomorphism
$F:x_{ij}\to x^{q}_{ij}$ ( $x_{ij}$: matrix elements, $q=p^{a}$, $a\in{\bf N}$ ) 
is applied and then yield the finite reductive group $SL(n,{\bf F}_{q})$.
While, in our case, we will consider, for example,
${\bf F}_{q}\otimes{\bf g}$ or ${\bf F}_{q}\otimes({\bf h}\oplus_{\alpha\in R}{\bf g}_{\alpha})$ ( ${\bf g}\in {\rm Lie}(G)$ ),
namely, a so-called Lie algebra lattice. 
From this aspect, the geometry of a set of discrete stationary points is closer to arithmetic geometry.

As we have discussed in the previous section, Lie$(SU(2))$ and the corresponding Heisenberg algebra define curves.
Due to $SU(2)\simeq SO(3)$, the Casimir element of Lie$(SU(2))$ corresponds to that of Lie$(SO(3))$,
i.e., $L^{2}=L^{2}_{x}+L^{2}_{y}+L^{2}_{z}$ as the magnitude of three-dimensional angular momentum.
Then we recognize that the Riemann zeta function of $SU(2)$ is expressed by $L^{2}$.

The general theory of Galois representation is as follows:
Let $G$ be a profinite group ( a typical example is a Galois group ),
let $R$ be a locally compact topological ring, and let $M$ be a finitely generated $R$-module.
Then one considers the following continuous homomorphism~[22,29,33,80],
\begin{eqnarray}
\rho: G \to {\rm Aut}_{R}(M).
\end{eqnarray}
This morphism is called as a linear representation of $G$.
In our case, the set of stationary points as fixed points of a Galois group gives an example of $M$.
Moreover, if the rank of $M$ is $n$ over $R$ ( $n=2$ in the case of elliptic curve ), then
\begin{eqnarray}
\rho: G \to {\rm Aut}_{R}(M) \simeq GL(n,R) = \{g\in M_{n}(R)|\det(g)\in R^{\times}\}
\end{eqnarray}
is obtained. 
From this aspect, an automorphism of the set of stationary points of the space of the NG bosonic coordinates gives a possibility 
toward a Galois representation theory.
It is possible to choose $U(n)$, $O(n)$ or $Sp(n)$ as $GL(n,R)$ by adopting an appropriate condition ( algebraic structure ) in $M$.
All of the notions of decomposition group, inertia group, Frobenius morphism, unramified/ramified, ..., 
consider corresponding invariant sets under their group actions~[22,29,33,80].
For example, the cyclotomic extension of $U(1)$ case has an isomorphism with a finite field ${\bf F}_{q}$.
Thus, those tools of Galois representations and Galois cohomology ( hence, class field theory ) 
will be introduced into the framework of our NG theorem.
This can be understood by the fact that a Galois theory studies a symmetry of a number field.  
In practice in number theory, usually one has to introduce a geometric object such as elliptic curves or Abelian varieties,
and an examination of a geometric object by the method of \'{e}tale cohomology gives a concrete example of a Galois representation,
especially an $l$-adic representation ( so-called $p\ne l$ case )~[22,29,33,53,80].
In our NG theorem, a set of stationary points ( vacua ) corresponds to a geometric object in the above prescription: 
A set of stationary points give a cyclotomic extension, and it is an example of Abelian extension due to the Kronecker-Weber theorem~[54,85],
then we yield the $n=1$ case of (148).

Our perspective is summarized in the following diagram:
\begin{flushleft}
normal/generalized/anomalous NG theorem $\to$ 
class field theory, adele, idele $\to$ Langrands correspondence.
\end{flushleft}

\section{Concluding Remarks}

In conclusion, we have studied the mechanism, the counting law of the number of true NG bosons, geometric and number theoretical aspects,
of the anomalous NG theorem.
We have established the counting law of true NG bosons of the diagonal breaking scheme of the anomalous NG theorem from several approaches,
while a more general case remains as an open question:
Probably, from our several observations in this paper, 
it seems not easy to give a general formula/law of generic breaking schema in the anomalous NG theorem. 
Namely, it seems the case that there is no universal counting law which is 
always valid to any type of symmetry breaking scheme of the anomalous NG theorem.
While, our several results of formalisms, geometry of Lie algebras and Lie groups in the anomalous NG theorem, 
Lagrangian and the effective potential show their universality.

We have presented a generic Lagrangian which has a Lorentz-violating parameter,
which gives our anomalous NG theorem.
Kostelecky et al. study on Lorentz and CPT violations intensively, as a fundamental physics,
especially from the context of neutrino phenomenology~[18,47].
It is interesting for us to find some applications of our result in theory of Lorentz/CPT violations.
In a Poincar\'{e} invariant theory, Lorentz and CPT symmetries are deeply related with each other.
Thus, our anomalous NG theorem would be restricted by CPT symmetries to apply it to several examples.

We have another interesting issue we will consider in the next step.
In several well-known substances ( metals ), some quantum fluctuations they may be described as NG modes 
still survive temperature regions over $T_{c}$.
Our anomalous NG theorem could be applied to such situations with giving new aspects to understand a mechanism of ordering in substances. 
To describe such physically/experimentally observed situations, we can utilize several mathematical and physical methods
such as the Maurer-Cartan form and Cartan geometry, the Stone-von Neumann theorem and Heisenberg manifolds/groups/algebras,
submanifold geometry and topology ( since we have found the fact that there is an interaction between a submanifold and its complement ),
quantum uncertainties, quantum fluctuations and quantum critical phenomena in quantum phase transitions.

Now we have arrived at the stage to modify/improve the traditional statement of the NG theorem in nonrelativistic/Lorentz-violated systems, 
given usually in several literatures, such as the paper of F. Strocchi~[79].
The first modification is to the usual statement that it argues the one-to-one correspondence between 
broken generators and the NG bosons with vanishing masses. 
In our case, the space of broken generators is "reduced", i.e., projected into a space of smaller dimensions.
From our result given in this paper, the notion of symmetry breaking is formulated as the following formal statement ( see the theorem given below ).
Usually, one employs the formalism of axiomatic field theory in literature, namely, 
(a) the definition of local quantum field theory,
(b) the definition of spontaneous symmetry breaking ,
(c) the nonrelativistic NG theorem,
(d) the relativistic NG theorem, in which some restrictions is applied to the nonrelativistic formalism,
especially due to the definition of conserved charge: 
In a nonrelativistic case, a three-dimensional support is used to define the integration domain of a N\"{o}ther current,
while a four-dimensional support will be prepared in a relativistic case.
Beside those delicate questions, in our formal statement, 
we do not need to distinguish relativistic and nonrelativistic cases seriously since the essential mechanism of anomalous NG theorem 
is the same between them.

The definition of spontaneously broken symmetry is improved to be:
\begin{flushleft}
{\bf Theorem}: 
Let $\beta$ be an internal symmetry described as a Lie group automorphism of an algebra $\cal{A}$,
which commutes with any spacetime translations.
After a symmetry breaking takes place, $\beta$ is fall into a representation $A\in \pi(\cal{A})$
and $\langle \beta{A}\rangle \ne \langle A\rangle$ modulo a lattice of Lie group,
and $\langle \beta{A}\rangle = \langle A\rangle$ when $\beta$ coincides with a lattice.
\end{flushleft}

\end{document}